\documentclass[conference]{IEEEtran}
\IEEEoverridecommandlockouts
\usepackage{cite}
\usepackage{amsmath,amssymb,amsfonts}
\DeclareMathOperator*{\argmax}{arg\,max}

\usepackage{enumitem}
\usepackage{bm}
\usepackage{graphicx}
\usepackage{mathrsfs}
\usepackage{amsthm}
\usepackage{algorithmic}
\usepackage{graphicx}
\usepackage{textcomp}
\usepackage{xcolor}
\usepackage{glossaries}
\newacronym{3gpp}{3GPP}{3rd Generation Partnership Project}
\newacronym{adc}{ADC}{Analog to Digital Converter}
\newacronym{5g}{5G}{5th generation}
\newacronym{6g}{6G}{6th generation}
\newacronym{aimd}{AIMD}{Additive Increase Multiplicative Decrease}
\newacronym{am}{AM}{Acknowledged Mode}
\newacronym{amc}{AMC}{Adaptive Modulation and Coding}
\newacronym{aqm}{AQM}{Active Queue Management}
\newacronym{awgn}{AGWN}{Additive White Gaussian Noise}
\newacronym{balia}{BALIA}{Balanced Link Adaptation}
\newacronym{bdp}{BDP}{Bandwidth-Delay Product}
\newacronym{bf}{BF}{beamforming}
\newacronym{cc}{CC}{Congestion Control}
\newacronym{cdf}{CDF}{Cumulative Distribution Function}
\newacronym{cn}{CN}{Core Network}
\newacronym{cqi}{CQI}{Channel Quality Information}
\newacronym{cp}{CP}{Control Plane}
\newacronym{csirs}{CSI-RS}{Channel State Information - Reference Signal}
\newacronym{df}{DF}{decode-and-forward}
\newacronym{rb}{RB}{Resource Block}
\newacronym{dce}{DCE}{Direct Code Execution}
\newacronym{dci}{DCI}{Downlink Control Information}
\newacronym{udp}{UDP}{User Datagram Protocol}
\newacronym{dl}{DL}{Downlink}
\newacronym{dmr}{DMR}{Deadline Miss Ratio}
\newacronym{dmrs}{DMRS}{DeModulation Reference Signal}
\newacronym{e2e}{E2E}{End-to-End}
\newacronym{ppp}{PPP}{Poission Point Process}
\newacronym{si}{SI}{Study Item}
\newacronym{ecn}{ECN}{Explicit Congestion Notification}
\newacronym{edf}{EDF}{Earliest Deadline First}
\newacronym{enb}{eNB}{eNodeB}
\newacronym{epc}{EPC}{Evolved Packet Core}
\newacronym{es}{ES}{Edge Server}
\newacronym{cav}{CAV}{Connected and Autonomous Vehicle}
\newacronym{fdma}{FDMA}{Frequency Division Multiple Access}
\newacronym{fdd}{FDD}{Frequency Division Duplexing}
\newacronym{upa}{UPA}{Uniform Planar Array}
\newacronym[firstplural=Radio Access Technologies (RATs)]{rat}{RAT}{Radio Access Technology}
\newacronym[firstplural=Radio Access Technology (RTs)]{rt}{RT}{Radio Technology}
\newacronym{fs}{FS}{Fast Switching}
\newacronym{isd}{ISD}{inter-site distance}
\newacronym{ftp}{FTP}{File Transfer Protocol}
\newacronym{gnb}{gNB}{Next Generation Node Base}
\newacronym{harq}{HARQ}{Hybrid Automatic Repeat reQuest}
\newacronym{hetnet}{HetNet}{Heterogeneous Network}
\newacronym{hh}{HH}{Hard Handover}
\newacronym{hol}{HOL}{Head-of-Line}
\newacronym{ia}{IA}{Initial Access}
\newacronym{imt}{IMT}{International Mobile Telecommunication}
\newacronym{iot}{IoT}{Internet of Things}
\newacronym{miot}{mIoT}{massive Internet of Things}
\newacronym{los}{LOS}{Line of Sight}
\newacronym{lte}{LTE}{Long Term Evolution}
\newacronym{m2m}{M2M}{Machine to Machine}
\newacronym{mac}{MAC}{Medium Access Control}
\newacronym{mc}{MC}{Multi-Connectivity}
\newacronym{mcs}{MCS}{Modulation and Coding Scheme}
\newacronym{mec}{MEC}{Mobile Edge Cloud}
\newacronym{mi}{MI}{Mutual Information}
\newacronym{mimo}{MIMO}{Multiple Input Multiple Output}
\newacronym{mmwave}{mmWave}{millimeter wave}
\newacronym{mptcp}{MPTCP}{Multipath TCP}
\newacronym{mr}{MR}{Maximum Rate}
\newacronym{mss}{MSS}{Maximum Segment Size}
\newacronym{mtd}{MTD}{Machine-Type Device}
\newacronym{mtu}{MTU}{Maximum Transmission Unit}
\newacronym{nfv}{NFV}{Network Function Virtualization}
\newacronym{vnf}{VNF}{ Virtualization Network Function}
\newacronym{sdn}{SDN}{Software Defined Networking}
\newacronym{nlos}{NLOS}{Non Line of Sight}
\newacronym{nlosb}{NLOSb}{Building Non Line of Sight}
\newacronym{nlosv}{NLOSv}{Vehicle Non Line of Sight}
\newacronym{nr}{NR}{New Radio}
\newacronym{ofdm}{OFDM}{Orthogonal Frequency Division Multiplexing}
\newacronym{ofdma}{OFDMA}{Orthogonal Frequency Division Multiple Access}
\newacronym{pdcch}{PDCCH}{Physical Downlonk Control Channel}
\newacronym{pdcp}{PDCP}{Packet Data Convergence Protocol}
\newacronym{pdsch}{PDSCH}{Physical Downlink Shared Channel}
\newacronym{pdu}{PDU}{Packet Data Unit}
\newacronym{pf}{PF}{Proportional Fair}
\newacronym{pgw}{PGW}{Packet Gateway}
\newacronym{phy}{PHY}{Physical}
\newacronym{pbch}{PBCH}{Physical Broadcast Channel}
\newacronym[plural=\gls{mme}s,firstplural=Mobility Management Entities (MMEs)]{mme}{MME}{Mobility Management Entity}
\newacronym{prb}{PRB}{Physical Resource Block}
\newacronym{pss}{PSS}{Primary Synchronization Signal}
\newacronym{pucch}{PUCCH}{Physical Uplink Control Channel}
\newacronym{pusch}{PUSCH}{Physical Uplink Shared Channel}
\newacronym{rach}{RACH}{Random Access Channel}
\newacronym{ran}{RAN}{Radio Access Network}
\newacronym{red}{RED}{Random Early Detection}
\newacronym{rf}{RF}{Radio Frequency}
\newacronym{rlc}{RLC}{Radio Link Control}
\newacronym{rlf}{RLF}{Radio Link Failure}
\newacronym{rrc}{RRC}{Radio Resource Control}
\newacronym{rrm}{RRM}{Radio Resource Management}
\newacronym{rr}{RR}{Round Robin}
\newacronym{rs}{RS}{Remote Server}
\newacronym{rsrp}{RSRP}{Reference Signal Received Power}
\newacronym{rss}{RSS}{Received Signal Strength}
\newacronym{rtt}{RTT}{Round Trip Time}
\newacronym{rw}{RW}{Receive Window}
\newacronym{rx}{RX}{Receiver}
\newacronym{sa}{SA}{standalone}
\newacronym{sack}{SACK}{Selective Acknowledgment}
\newacronym{sap}{SAP}{Service Access Point}
\newacronym{sch}{SCH}{Secondary Cell Handover}
\newacronym{scoot}{SCOOT}{Split Cycle Offset Optimization Technique}
\newacronym{sdma}{SDMA}{Spatial Division Multiple Access}
\newacronym{sinr}{SINR}{Signal to Interference plus Noise Ratio}
\newacronym{sm}{SM}{Saturation Mode}
\newacronym{snr}{SNR}{Signal to Noise Ratio}
\newacronym{son}{SON}{Self-Organizing Network}
\newacronym{ss}{SS}{Synchronization Signal}
\newacronym{srs}{SRS}{Sounding Reference Signal}
\newacronym{sss}{SSS}{Secondary Synchronization Signal}
\newacronym{tb}{TB}{Transport Block}
\newacronym{tcp}{TCP}{Transmission Control Protocol}
\newacronym{tdd}{TDD}{Time Division Duplexing}
\newacronym{tdma}{TDMA}{Time Division Multiple Access}
\newacronym{tfl}{TfL}{Transport for London}
\newacronym{tm}{TM}{Transparent Mode}
\newacronym{prr}{PRR}{Packet Reception Ratio}
\newacronym{trp}{TRP}{Transmitter Receiver Pair}
\newacronym{tti}{TTI}{Transmission Time Interval}
\newacronym{ttt}{TTT}{Time-to-Trigger}
\newacronym{tx}{TX}{Transmitter}
\newacronym{ue}{UE}{User Equipment}
\newacronym{ul}{UL}{Uplink}
\newacronym{uml}{UML}{Unified Modeling Language}
\newacronym{um}{UM}{Unacknowledged Mode}
\newacronym{utc}{UTC}{Urban Traffic Control}
\newacronym{vm}{VM}{Virtual Machine}
\newacronym{rsrq}{RSRQ}{Reference Signal Received Quality}
\newacronym{rssi}{RSSI}{Received Signal Strength Indicator}
\newacronym{crs}{CRS}{Cell Reference Signal}
\newacronym{v2v}{V2V}{Vehicle-to-Vehicle}
\newacronym{v2i}{V2I}{Vehicle-to-Infrastructure}
\newacronym{v2n}{V2N}{Vehicle-to-Network}
\newacronym{v2x}{V2X}{Vehicle-to-Everything}
\newacronym{vn}{VN}{Vehicular Node}
\newacronym{dsrc}{DSRC}{Dedicated Short Range Communication}
\newacronym{ci}{CI}{context information}
\newacronym{voi}{VoI}{value of information}
\newacronym{gps}{GPS}{Global Positioning System}
\newacronym{qos}{QoS}{Quality of Service}
\newacronym{qoe}{QoE}{Quality of Experience}
\newacronym{ml}{ML}{Machine Learning}
\newacronym{ahp}{AHP}{Analytic Hierarchy Process}
\newacronym{lidar}{LIDAR}{Light Detection and Ranging}
\newacronym{sumo}{SUMO}{Simulation of Urban MObility}
\newacronym{wave}{WAVE}{Wireless Access in Vehicular Environment}
\newacronym{c-its}{C-ITS}{Connected Intelligent Transportation System}
\newacronym{dash}{DASH}{Dynamic Adaptive Streaming over HTTP}
\newacronym{http}{HTTP}{HyperText Transfer Protocol}
\newacronym{nt}{NT}{non-terrestrial}
\newacronym{ntc}{NTC}{non-terrestrial communication}
\newacronym{ntn}{NTN}{non-terrestrial network}
\newacronym{haps}{HAPS}{High Altitude Platform Station}
\newacronym{hap}{HAP}{High Altitude Platform}
\newacronym{lap}{LAP}{Low Altitude Platform}
\newacronym{leo}{LEO}{Low Earth Orbit}
\newacronym{meo}{MEO}{Medium Earth Orbit}
\newacronym{geo}{GEO}{Geostationary Earth Orbit}
\newacronym{uav}{UAV}{Unmanned Aerial Vehicle}
\newacronym{nsat}{nSAT}{Nanosatellite}
\newacronym{ehf}{EHF}{extremely high-frequency}
\newacronym{ioe}{IoE}{Internet of Everyone}
\newacronym{gan}{GaN}{Gallium Nitride}
\newacronym{nbiot}{NB-IoT}{Narrowband-IoT}
\newacronym{lora}{LoRa}{Long Range}
\newacronym{ism}{ISM}{Industrial, Scientific and Medical}
\newacronym{lpwan}{LPWAN}{Low-Power Wide Area Network}
\newacronym{css}{CSS}{Chirp Spread Spectrum}
\newacronym{sf}{SF}{Spreading Factor}
\newacronym{qpsk}{QPSK}{Quadrature Phase Shift Keying}
\newacronym{mar}{MAR}{Mobile Autonomous Reporting}
\newacronym{aoi}{AoI}{area of interest} 
\newacronym{toa}{ToA}{time on air}
\newacronym{fhss}{FH-SS}{frequency-hopping spread-spectrum}
\usepackage{hyperref}
\usepackage[nolist]{acronym}
\usepackage[capitalize]{cleveref}
\crefname{section}{Sec.}{Secs.}

\usepackage[font=scriptsize]{subcaption}
\usepackage[font=footnotesize]{caption}

\usepackage{tikz}
\usepackage{pgfplots}
\usepackage{tikzscale}
\usepackage{tikz-qtree}

\pgfplotsset{compat=newest}
\pgfplotsset{plot coordinates/math parser=false}
\pgfplotsset{every axis/.append style={
                    label style={font=\scriptsize},
                    tick label style={font=\scriptsize},
                    legend style={font=\scriptsize}
                    }}
\usetikzlibrary{plotmarks,shapes,patterns,decorations.pathreplacing,backgrounds,calc,arrows,arrows.meta,spy,matrix,shadows,trees,positioning,fit}
\usepgfplotslibrary{patchplots,groupplots}

\tikzstyle{startstop} = [rectangle, rounded corners, minimum width=2cm, minimum height=0.5cm,text centered, draw=black]
\tikzstyle{io} = [trapezium, trapezium left angle=70, trapezium right angle=110, minimum width=3cm, minimum height=1cm, text centered, draw=black]
\tikzstyle{process} = [rectangle, minimum width=2cm, minimum height=0.5cm, text centered, draw=black, alignb=center]
\tikzstyle{decision} = [ellipse, minimum width=2cm, minimum height=1cm, text centered, draw=black]
\tikzstyle{arrow} = [thick,<->,>=stealth]
\tikzstyle{line} = [thick,>=stealth]
\tikzstyle{darrow} = [thick,<->,>=stealth,dashed]
\tikzstyle{sarrow} = [thick,->,>=stealth]
\tikzstyle{larrow} = [line width=0.1mm,dashdotted,->,>=stealth]

\pgfkeys{/pgf/number format/.cd,1000 sep={\,}} 

\makeatletter
\def\grd@save@target#1{%
  \def\grd@target{#1}}
\def\grd@save@start#1{%
  \def\grd@start{#1}}
\tikzset{
  grid with coordinates/.style={
    to path={%
      \pgfextra{%
        \edef\grd@@target{(\tikztotarget)}%
        \tikz@scan@one@point\grd@save@target\grd@@target\relax
        \edef\grd@@start{(\tikztostart)}%
        \tikz@scan@one@point\grd@save@start\grd@@start\relax
        \draw[minor help lines] (\tikztostart) grid (\tikztotarget);
        \draw[major help lines] (\tikztostart) grid (\tikztotarget);
        \grd@start
        \pgfmathsetmacro{\grd@xa}{\the\pgf@x/1cm}
        \pgfmathsetmacro{\grd@ya}{\the\pgf@y/1cm}
        \grd@target
        \pgfmathsetmacro{\grd@xb}{\the\pgf@x/1cm}
        \pgfmathsetmacro{\grd@yb}{\the\pgf@y/1cm}
        \pgfmathsetmacro{\grd@xc}{\grd@xa + \pgfkeysvalueof{/tikz/grid with coordinates/major step x}}
        \pgfmathsetmacro{\grd@yc}{\grd@ya + \pgfkeysvalueof{/tikz/grid with coordinates/major step y}}
        \foreach \x in {\grd@xa,\grd@xc,...,\grd@xb}
        \node[anchor=north] at (\x,\grd@ya) {\pgfmathprintnumber{\x}};
        \foreach \y in {\grd@ya,\grd@yc,...,\grd@yb}
        \node[anchor=east] at (\grd@xa,\y) {\pgfmathprintnumber{\y}};
      }
    }
  },
  minor help lines/.style={
    help lines,
    gray,
    line cap =round,
    xstep=\pgfkeysvalueof{/tikz/grid with coordinates/minor step x},
    ystep=\pgfkeysvalueof{/tikz/grid with coordinates/minor step y}
  },
  major help lines/.style={
    help lines,
    line cap =round,
    line width=\pgfkeysvalueof{/tikz/grid with coordinates/major line width},
    xstep=\pgfkeysvalueof{/tikz/grid with coordinates/major step x},
    ystep=\pgfkeysvalueof{/tikz/grid with coordinates/major step y}
  },
  grid with coordinates/.cd,
  minor step x/.initial=.5,
  minor step y/.initial=.2,
  major step x/.initial=1,
  major step y/.initial=1,
  major line width/.initial=1pt,
}
\makeatother

\newlength\fheight
\newlength\fwidth
\usetikzlibrary{patterns}
\pgfplotsset{compat=newest}
\pgfplotsset{plot coordinates/math parser=false}
\usetikzlibrary{plotmarks,patterns,decorations.pathreplacing,backgrounds,calc,arrows,arrows.meta,spy,matrix}
\usepgfplotslibrary{patchplots,groupplots}
\usepackage{tikzscale}
\usepackage{multirow}

\def\BibTeX{{\rm B\kern-.05em{\sc i\kern-.025em b}\kern-.08em
    T\kern-.1667em\lower.7ex\hbox{E}\kern-.125emX}}
\begin{document}

\begin{acronym}

\acro{5G-NR}{5G New Radio}
\acro{3GPP}{3rd Generation Partnership Project}
\acro{AC}{address coding}
\acro{ACF}{autocorrelation function}
\acro{ACR}{autocorrelation receiver}
\acro{ADC}{analog-to-digital converter}
\acrodef{aic}[AIC]{Analog-to-Information Converter}     
\acro{AIC}[AIC]{Akaike information criterion}
\acro{aric}[ARIC]{asymmetric restricted isometry constant}
\acro{arip}[ARIP]{asymmetric restricted isometry property}

\acro{ARQ}{automatic repeat request}
\acro{AUB}{asymptotic union bound}
\acrodef{awgn}[AWGN]{Additive White Gaussian Noise}     
\acro{AWGN}{additive white Gaussian noise}
\acro{ToA}{Time on Air}
\acro{CER}{Code Error Rate}
\acro{DtS}{direct-to-satellite}
\acro{PHY}{physical layer}
\acro{FH}{frequency hopping}
\acro{UNB}{ultra narrowband}
\acro{DCSS}{Differential chirp spread spectrum}
\acro{GEO}{geosynchronous equatorial orbit}
\acro{CSS}{Chirp Spread Spectrum}
\acro{RFID}{Radio Frequency Identification}
\acro{LoS}{line of sight}
\acro{CRC}{cyclic redundancy check}
\acro{FEC}{forward error correction}
\acro{LPWAN}{low power wide area network}
\acro{HiGFC-FH}{High GF Coding Frequency Hopping}
\acro{MDS}{Maximum Distance Separable}
\acro{RTT}{Round Trip Time}
\acro{OCW}{Operating Channel Width}
\acro{OBW}{Occupied Band Width}
\acro{MSB}{minimum separation band}
\acro{FEC}{forward error correction}
\acro{TFA}{Time/Frequency Aloha}
\acro{LPGAN}{low-power global area network}
\acro{BCH}{Bose–Chaudhuri–Hocquenghem codes}
\acro{PA}{Pure Aloha}
\acro{RA}{Random Access}
\acro{LR-FHSS}{Long Range - Frequency Hopping Spread Specturm}
\acro{HLC-FH}{Hop-level Coding Frequency hopping}

\acro{APSK}[PSK]{asymmetric PSK} 

\acro{waric}[AWRICs]{asymmetric weak restricted isometry constants}
\acro{warip}[AWRIP]{asymmetric weak restricted isometry property}
\acro{BCH}{Bose, Chaudhuri, and Hocquenghem}        
\acro{BCHC}[BCHSC]{BCH based source coding}
\acro{BEP}{bit error probability}
\acro{BFC}{block fading channel}
\acro{BG}[BG]{Bernoulli-Gaussian}
\acro{BGG}{Bernoulli-Generalized Gaussian}
\acro{BPAM}{binary pulse amplitude modulation}
\acro{BPDN}{Basis Pursuit Denoising}
\acro{BPPM}{binary pulse position modulation}
\acro{BPSK}{binary phase shift keying}
\acro{BPZF}{bandpass zonal filter}
\acro{BSC}{binary symmetric channels}              
\acro{BU}[BU]{Bernoulli-uniform}
\acro{BER}{bit error rate}
\acro{BS}{base station}

\acro{CP}{Cyclic Prefix}
\acrodef{cdf}[CDF]{cumulative distribution function}   
\acro{CDF}{cumulative distribution function}
\acrodef{c.d.f.}[CDF]{cumulative distribution function}
\acro{CCDF}{complementary cumulative distribution function}
\acrodef{ccdf}[CCDF]{complementary CDF}               
\acrodef{c.c.d.f.}[CCDF]{complementary cumulative distribution function}
\acro{CD}{cooperative diversity}

\acro{CDMA}{Code Division Multiple Access}
\acro{ch.f.}{characteristic function}
\acro{CIR}{channel impulse response}
\acro{cosamp}[CoSaMP]{compressive sampling matching pursuit}
\acro{CR}{cognitive radio}
\acro{cs}[CS]{compressed sensing}                   
\acrodef{cscapital}[CS]{Compressed sensing} 
\acrodef{CS}[CS]{compressed sensing}
\acro{CSI}{channel state information}
\acro{CCSDS}{consultative committee for space data systems}
\acro{CC}{convolutional coding}
\acro{Covid19}[COVID-19]{Coronavirus disease}

\acro{DAA}{detect and avoid}
\acro{DAB}{digital audio broadcasting}
\acro{DCT}{discrete cosine transform}
\acro{dft}[DFT]{discrete Fourier transform}
\acro{DR}{distortion-rate}
\acro{DS}{direct sequence}
\acro{DS-SS}{direct-sequence spread-spectrum}
\acro{DTR}{differential transmitted-reference}
\acro{DVB-H}{digital video broadcasting\,--\,handheld}
\acro{DVB-T}{digital video broadcasting\,--\,terrestrial}
\acro{DL}{downlink}
\acro{DSSS}{Direct Sequence Spread Spectrum}
\acro{DFT-s-OFDM}{Discrete Fourier Transform-spread-Orthogonal Frequency Division Multiplexing}
\acro{DAS}{distributed antenna system}
\acro{DNA}{Deoxyribonucleic Acid}
\acro{DF}{decode-and-forward}
\acro{DBPSK}{Differential Binary Phase Shift Keying}

\acro{EC}{European Commission}
\acro{EED}[EED]{exact eigenvalues distribution}
\acro{EIRP}{Equivalent Isotropically Radiated Power}
\acro{ELP}{equivalent low-pass}
\acro{eMBB}{Enhanced Mobile Broadband}
\acro{EMF}{electric and magnetic fields}
\acro{EU}{European union}

\acro{FC}[FC]{fusion center}
\acro{FCC}{Federal Communications Commission}
\acro{FEC}{forward error correction}
\acro{FFT}{fast Fourier transform}
\acro{FH}{frequency-hopping}
\acro{FH-SS}{frequency-hopping spread-spectrum}
\acrodef{FS}{Frame synchronization}
\acro{FSsmall}[FS]{frame synchronization}  
\acro{FDMA}{Frequency Division Multiple Access}

\acro{GA}{Gaussian approximation}
\acro{GF}{Galois field }
\acro{GG}{Generalized-Gaussian}
\acro{GIC}[GIC]{generalized information criterion}
\acro{GLRT}{generalized likelihood ratio test}
\acro{GPS}{Global Positioning System}
\acro{GMSK}{Gaussian minimum shift keying}
\acro{GSMA}{Global System for Mobile communications Association}
\acro{GS}{ground station}
\acro{HAP}{high altitude platform}

\acro{IDR}{information distortion-rate}
\acro{IFFT}{inverse fast Fourier transform}
\acro{iht}[IHT]{iterative hard thresholding}
\acro{i.i.d.}{independent, identically distributed}
\acro{IoT}{Internet of Things}                      
\acro{IR}{impulse radio}
\acro{lric}[LRIC]{lower restricted isometry constant}
\acro{lrict}[LRICt]{lower restricted isometry constant threshold}
\acro{ISI}{intersymbol interference}
\acro{ITU}{International Telecommunication Union}
\acro{ICNIRP}{International Commission on Non-Ionizing Radiation Protection}
\acro{IEEE}{Institute of Electrical and Electronics Engineers}
\acro{ICES}{IEEE international committee on electromagnetic safety}
\acro{IEC}{International Electrotechnical Commission}
\acro{IARC}{International Agency on Research on Cancer}
\acro{IS-95}{Interim Standard 95}
\acro{ISM}{Industrial, Scientific, and Medical}

\acro{km}{kilometer}

\acro{LEO}{low earth orbit}
\acro{LF}{likelihood function}
\acro{LLF}{log-likelihood function}
\acro{LLR}{log-likelihood ratio}
\acro{LLRT}{log-likelihood ratio test}
\acro{LOS}{Line-of-Sight}
\acro{LRT}{likelihood ratio test}
\acro{wlric}[LWRIC]{lower weak restricted isometry constant}
\acro{wlrict}[LWRICt]{LWRIC threshold}
\acro{LoRaWAN}{Low power long Range Wide Area Network}
\acro{NLOS}{non-line-of-sight}
\acro{LAP}{low altitude platform}

\acro{MB}{multiband}
\acro{MC}{multicarrier}
\acro{MDS}{mixed distributed source}
\acro{MF}{matched filter}
\acro{m.g.f.}{moment generating function}
\acro{MI}{mutual information}
\acro{MIMO}{multiple-input multiple-output}
\acro{MISO}{multiple-input single-output}
\acrodef{maxs}[MJSO]{maximum joint support cardinality}                       
\acro{ML}[ML]{maximum likelihood}
\acro{MMSE}{minimum mean-square error}
\acro{MMV}{multiple measurement vectors}
\acrodef{MOS}{model order selection}
\acro{M-PSK}[${M}$-PSK]{$M$-ary phase shift keying}                       
\acro{M-APSK}[${M}$-PSK]{$M$-ary asymmetric PSK} 
\acro{MSD}{Maximum Serving Distance}

\acro{M-QAM}[$M$-QAM]{$M$-ary quadrature amplitude modulation}
\acro{MRC}{maximal ratio combiner}                  
\acro{maxs}[MSO]{maximum sparsity order}                                      
\acro{M2M}{machine to machine}                                                
\acro{MUI}{multi-user interference}
\acro{mMTC}{massive Machine Type Communications}      
\acro{mm-Wave}{millimeter-wave}
\acro{MP}{mobile phone}
\acro{MPE}{maximum permissible exposure}
\acro{MAC}{media access control}
\acro{NB}{narrowband}
\acro{NBI}{narrowband interference}
\acro{NLA}{nonlinear sparse approximation}
\acro{NLOS}{Non-Line of Sight}
\acro{NTIA}{National Telecommunications and Information Administration}
\acro{NTP}{National Toxicology Program}
\acro{NHS}{National Health Service}
\acro{NTN}{Non-terrestrial Networks}

\acro{OC}{optimum combining}                             
\acro{OC}{optimum combining}
\acro{ODE}{operational distortion-energy}
\acro{ODR}{operational distortion-rate}
\acro{OFDM}{orthogonal frequency-division multiplexing}
\acro{omp}[OMP]{orthogonal matching pursuit}
\acro{OSMP}[OSMP]{orthogonal subspace matching pursuit}
\acro{OQAM}{offset quadrature amplitude modulation}
\acro{OQPSK}{offset QPSK}
\acro{OFDMA}{Orthogonal Frequency-division Multiple Access}
\acro{OPEX}{Operating Expenditures}
\acro{OQPSK/PM}{OQPSK with phase modulation}

\acro{PAM}{pulse amplitude modulation}
\acro{PAR}{peak-to-average ratio}
\acrodef{pdf}[PDF]{probability density function}                      
\acro{PDF}{probability density function}
\acrodef{p.d.f.}[PDF]{probability distribution function}
\acro{PDP}{power dispersion profile}
\acro{PMF}{probability mass function}                             
\acrodef{p.m.f.}[PMF]{probability mass function}
\acro{PN}{pseudo-noise}
\acro{PPM}{pulse position modulation}
\acro{PRake}{Partial Rake}
\acro{PSD}{power spectral density}
\acro{PSEP}{pairwise synchronization error probability}
\acro{PSK}{phase shift keying}
\acro{PD}{power density}
\acro{8-PSK}[$8$-PSK]{$8$-phase shift keying}
\acro{PS}{the probability of successful transmission}
 
\acro{FSK}{frequency shift keying}
\acro{GFSK}{Gaussian Frequency-Shift Keying}

\acro{QAM}{Quadrature Amplitude Modulation}
\acro{QPSK}{quadrature phase shift keying}
\acro{OQPSK/PM}{OQPSK with phase modulator }

\acro{RD}[RD]{raw data}
\acro{RDL}{"random data limit"}
\acro{ric}[RIC]{restricted isometry constant}
\acro{rict}[RICt]{restricted isometry constant threshold}
\acro{rip}[RIP]{restricted isometry property}
\acro{ROC}{receiver operating characteristic}
\acro{rq}[RQ]{Raleigh quotient}
\acro{RS}[RS]{Reed-Solomon}
\acro{RSC}[RSSC]{RS based source coding}
\acro{r.v.}{random variable}                               
\acro{R.V.}{random vector}
\acro{RMS}{root mean square}
\acro{RFR}{radiofrequency radiation}
\acro{RIS}{Reconfigurable Intelligent Surface}
\acro{RNA}{RiboNucleic Acid}

\acro{S}{Sensitivity}
\acro{SA}[SA-Music]{subspace-augmented MUSIC with OSMP}
\acro{SCBSES}[SCBSES]{Source Compression Based Syndrome Encoding Scheme}
\acro{SCM}{sample covariance matrix}
\acro{SEP}{symbol error probability}
\acro{SG}[SG]{sparse-land Gaussian model}
\acro{SIMO}{single-input multiple-output}
\acro{SINR}{signal-to-interference plus noise ratio}
\acro{SIR}{signal-to-interference ratio}
\acro{SISO}{single-input single-output}
\acro{SMV}{single measurement vector}
\acro{SNR}[\textrm{SNR}]{signal-to-noise ratio} 
\acro{sp}[SP]{subspace pursuit}
\acro{SS}{spread spectrum}
\acro{SW}{sync word}
\acro{SAR}{specific absorption rate}
\acro{SSB}{synchronization signal block}
\acro{SF}{Spreading Factor}
\acro{MAR}{Mobile Autonomous Reporting}

\acro{TH}{time-hopping}
\acro{TR}{transmitted-reference}
\acro{TW}{Tracy-Widom}
\acro{TWDT}{TW Distribution Tail}
\acro{TCM}{trellis coded modulation}
\acro{TDD}{time-division duplexing}
\acro{TDMA}{Time Division Multiple Access}

\acro{UAV}{unmanned aerial vehicle}
\acro{uric}[URIC]{upper restricted isometry constant}
\acro{urict}[URICt]{upper restricted isometry constant threshold}
\acro{UWB}{ultrawide band}
\acro{UWBcap}[UWB]{Ultrawide band}   
\acro{URLLC}{Ultra Reliable Low Latency Communications}
         
\acro{wuric}[UWRIC]{upper weak restricted isometry constant}
\acro{wurict}[UWRICt]{UWRIC threshold}                
\acro{UE}{user equipment}
\acro{UL}{uplink}
\acro{UNB}{Ultra-NarrowBand}

\acro{WiM}[WiM]{weigh-in-motion}
\acro{WLAN}{wireless local area network}
\acro{wm}[WM]{Wishart matrix}                               
\acroplural{wm}[WM]{Wishart matrices}
\acro{WMAN}{wireless metropolitan area network}
\acro{WPAN}{wireless personal area network}
\acro{wric}[WRIC]{weak restricted isometry constant}
\acro{wrict}[WRICt]{weak restricted isometry constant thresholds}
\acro{wrip}[WRIP]{weak restricted isometry property}
\acro{WSN}{wireless sensor network}                        
\acro{WSS}{wide-sense stationary}
\acro{WHO}{World Health Organization}
\acro{Wi-Fi}{wireless fidelity}

\acro{sss}[SpaSoSEnc]{sparse source syndrome encoding}

\acro{VLC}{visible light communication}
\acro{VPN}{virtual private network} 
\acro{RF}{radio frequency}
\acro{FSO}{free space optics}
\acro{IoST}{Internet of space things}

\acro{GSM}{Global System for Mobile Communications}
\acro{2G}{second-generation cellular network}
\acro{3G}{third-generation cellular network}
\acro{4G}{fourth-generation cellular network}
\acro{5G}{5th-generation cellular network}	
\acro{6G}{6th-generation cellular network}
\acro{gNB}{next generation node B base station}
\acro{NR}{New Radio}
\acro{UMTS}{Universal Mobile Telecommunications Service}
\acro{LTE}{Long Term Evolution}

\acro{QoS}{Quality of Service}
\end{acronym}

\title{On the Performance of Non-Terrestrial Networks \\ to Support the Internet of Things}

    \author{\IEEEauthorblockN{Dengke Wang$^{\star }$, Alessandro Traspadini$^{\circ }$, Marco Giordani$^{\circ }$, Mohamed-Slim Alouini$^{\star }$, Michele Zorzi$^{\circ }$\medskip}
\IEEEauthorblockA{
$^{\star}$King Abdullah University of Science and Technology (KAUST), SA. Email: \texttt{\{dengke.wang\}@kaust.edu.sa}\\
$^{\circ}$University of Padova, Dept. of Information Engineering (DEI), Italy. Email: \texttt{\{name.surname\}@dei.unipd.it}}}
    
    \maketitle
    
    \begin{abstract}
    The advent of the \gls{iot} era, where billions of devices and sensors are becoming more and more connected and ubiquitous, is putting a strain on traditional terrestrial networks, that may no longer be able to fulfill service requirements efficiently. This issue is further complicated in rural and remote areas with scarce and low-quality cellular coverage. To fill this gap, the research community is focusing on \glspl{ntn}, where \glspl{uav}, \glspl{hap} and satellites can serve as aerial/space gateways to aggregate, process, and relay the IoT traffic. 
    In this paper we demonstrate this paradigm, and evaluate how common Low-Power Wide Area Network (LPWAN) technologies, designed and developed to operate for IoT systems, work in NTNs. We then formalize an optimization problem to decide whether and how IoT traffic can be offloaded to LEO satellites to reduce the burden on terrestrial gateways. 

    \end{abstract}

\begin{tikzpicture}[remember picture,overlay]
\node[anchor=north,yshift=-10pt] at (current page.north) {\parbox{\dimexpr\textwidth-\fboxsep-\fboxrule\relax}{
\centering\footnotesize This paper has been accepted for publication at the 56th Asilomar Conference on Signals, Systems, and Computers. \textcopyright 2022 IEEE.\\
Please cite it as: D. Wang, A. Traspadini, M. Giordani, M.-S. Alouini, and M. Zorzi, "On the Performance of Non-Terrestrial Networks
to Support the Internet of Things,"  56th Asilomar Conference on Signals, Systems, and Computers, 2022.}};
\end{tikzpicture}
    
\glsresetall

\section{Introduction} 
\label{sec:introduction}


\begin{figure*}[h!]
\centering 
\includegraphics[width=1\textwidth]{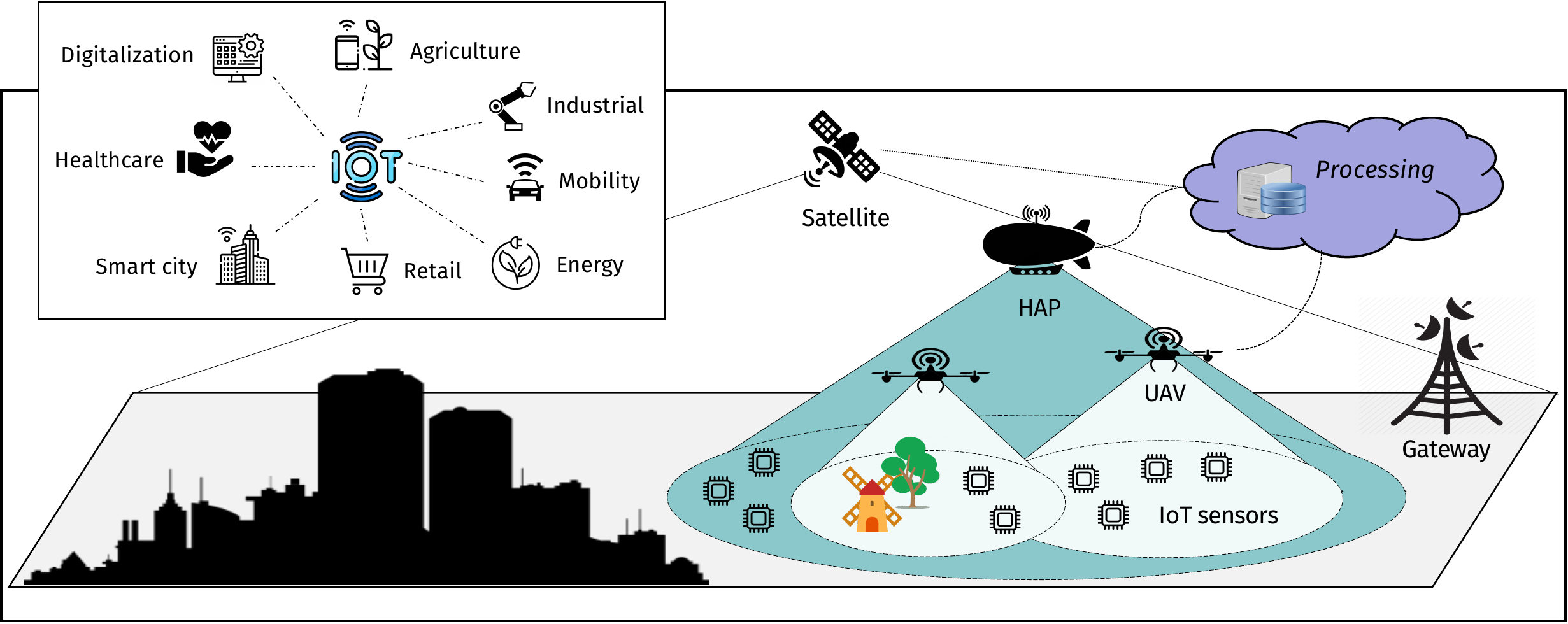}
\caption{The IoT-NTN scenario and relative use cases.}
\label{fig:iotntn}
\end{figure*}

By the end of 2025, the number of \gls{iot} devices will rise to 75 billions worldwide,
creating a global market of around 11.1 trillions USD according to some estimates~\cite{safaei2017reliability}.
Functional and robust IoT applications improve our life quality, and provide convenience in many fields, including transportation and logistics (e.g., to support assisted driving or help in the management of goods), healthcare (e.g., to improve workflow in hospitals or facilitate automatic data collection and sensing), agriculture (e.g., to monitor soil and crop parameters), and smart cities~\cite{atzori2010internet,zanella2014internet}.

In the 5G era, \gls{miot}, also known as massive Machine Type Communication (mMTC), promotes the support for extremely low-cost low-energy-consumption sensors (e.g., temperature, pressure, humidity, etc.)  that transmit small volumes of data but, cumulatively, generate large data rates.
To satisfy connectivity requests, standardization bodies and industry players have developed \gls{lpwan} technologies, such as \gls{lora}, \gls{nbiot}, and SigFox, which define different \gls{phy} and \gls{mac} layers, and operate in the sub-6 GHz~bands to provide good balance between range and performance~\cite{ayoub2018internet}.

Nevertheless, inter-connecting billions of smart devices may eventually congest traditional terrestrial networks which, at the same time, may be unable to serve end devices in rural/remote regions or in case of emergency where infrastructures are unavailable or out of order, respectively~\cite{chaoub20216g}.
To address these issues, the research community is exploring the concept of \glspl{ntn}~\cite{giordani2020non}, where \glspl{uav}, \glspl{hap} and satellites expand traditional two-dimensional networks by acting as aerial/space gateways operating from the sky, as illustrated in Fig.~\ref{fig:iotntn}.
Notably, these elements can provide very large continuous and autonomous geographical coverage, even in the absence of pre-existing terrestrial infrastructures, thus offering global connectivity for \gls{iot} applications that rely on sensors~\cite{vaezi2021cellular}. 
Potential beneficiaries of this paradigm, referred to as NTN-IoT, include inter-regional transport, unserved farmlands, ships, mountainous areas, and remote maintenance facilities.

In this context, while the literature generally focuses on IoT for smart cities (e.g.,~\cite{magrin2017performance,mroue2018mac}), some recent works have started to explore the applicability of \gls{lpwan} technologies to NTN-IoT scenarios.
However, most of the prior art considers standalone UAV~\cite{zhan2019energy}, HAP~\cite{ke2021edge}, or satellite~\cite{soret20205g} systems as a solution to gather and process IoT traffic from terrestrial networks, even though integrated/multilayered aerial/space architectures, as proposed in~\cite{wang2021potential}, may further improve quality of service.
Moreover, motivated by recent trends in the \gls{3gpp}, NTN-IoT deployments have been studied using \gls{nbiot}~\cite{liberg2020narrowband} and for satellite-only scenarios, with preliminary results published in~\cite{guidotti2020non}.
However, it is not clear whether some other \gls{lpwan} technologies, such as LoRa or SigFox, would provide superior performance for the same \gls{ntn}-IoT applications.
At the same time, the~literature often neglects needs and requirements of the rural environment, since most of the analysis is based on urban~scenarios.

To fill these gaps, in this paper we evaluate via simulation the performance of several \gls{ntn}-IoT configurations, considering different \gls{lpwan} technologies (i.e., LoRa, NB-IoT, SigFox) and non-terrestrial architectures (i.e., UAVs, HAPs, LEO satellites, and their combinations). Then, we provide guidelines on how to dimension these systems as a function of several parameters including the radius of the service area and the density of IoT sensors and gateways. 
We demonstrate that, while LoRa is the best option for LEO satellites in terms of both coverage and goodput, NB-IoT is more desirable to connect UAVs and HAPs.
Moreover, we raise the question of where to process IoT data, and develop an optimization problem to decide whether IoT sensors should offload (part of) their traffic to NTNs to reduce the congestion of terrestrial gateways.  
We see that the probability of successful transmission improves by up to 30\% when some processing tasks are delegated to LEO~satellites.

\section{Enabling Technologies for \\ \acrlong{lpwan}}
\label{sec:lpwan}
Vaezi \emph{et al.} introduce two main categories of use cases~\cite{vaezi2021cellular}, namely \acrfull{miot} and Critical Internet of Things (CIoT), where
mMTC is designed to support many low-cost sensors that continuously transmit small streams of data, and CIoT involves fewer devices handling larger volumes of data.
Industrial control, robotic machines, and autonomous vehicles are examples of CIoT, whereas mMTC describes applications for data collection through sensors, for example in smart agriculture and/or smart city scenarios~\cite{zhang2021ara}.
As such, the IoT market is fragmented, with many organizations promoting different (and somehow conflicting) access technologies and vertical solutions. 
In this work, we focus on mIoT, 
and compare three main \gls{lpwan} technologies, as described below and summarized in~\cref{tab:params}.

\begin{table*}[t!] \label{tab:params}
\renewcommand{\arraystretch}{1.7}
\caption{Summary of the LPWAN technologies. SF$k$, $k\in\{7,\dots,12\}$, is the \rm{SF} in LoRa, while $\mathcal{R}$ is the number of repetitions in NB-IoT.}
    \label{tab:params}
    \centering
\begin{tabular}{|l|c|c|c|}
\hline
{Characteristic} & Sigfox & LoRa & NB-IoT   \\ \hline
Modulation &  UNB  & CSS & QPSK \\ \hline
Bandwidth (Uplink) &  100 Hz &  125 kHz (Class A) & 180 kHz \\ \hline
Max. data rate (Uplink) [kbps] &  0.1 &  6 (SF7) & 90 (QPSK) \\ \hline
Max. range [km] &  50 & 14 (SF12) & 10 \\ \hline
Energy consumption &  Very low & Low & Low \\ \hline
Tx. power [dBm] &  14 & 14 & 23 \\ \hline
Interference immunity & FH-SS and repetition coding&  SF orthogonality & Repetition coding \\ \hline
Sensitivity threshold~[dBm] &  $-140$ dBm & $-127-2.5(\text{SF}k-7)$ &$-102.2-2.8\log_2(\mathcal{R})$  \\ \hline
Device cost~[USD] & 5 & 10 & 12 \\\hline
\end{tabular}
\end{table*}


\subsection{SigFox}

SigFox~\cite{ribeiro2018outdoor} devices operate in the 863/870-MHz ISM spectrum with a transmit power of 14 dBm in Europe. They use a bandwidth of 100 Hz (1.5 kHz) in uplink (downlink), offering a data rate of 100 (600) bps with 12 (8) bytes of maximum payload.
Given the small packet size, this solution promotes low energy consumption and prolonged battery life of the devices.
Using \ac{UNB} modulation, combined with \ac{DBPSK} and \ac{GFSK}, SigFox achieves wide-range communications between 10 and 50 km, and robustness against noise.
It exploits \gls{fhss} and  repetition code, where the transmitter copies the message into three slices and successively transmits them through three randomly selected sub-frequency bands, both of which provide immunity to interference.

\subsection{LoRa}
\label{sub:LoRa}

LoRa, including LoRaWAN, is a proprietary \gls{lpwan} technology designed and patented by Semtech~\cite{magrin2017performance}.
In this paper we consider LoRa Class A networks~\cite{centenaro2016long}, where transmissions are always initiated by the end devices. 
Specifically, LoRa devices operate in the 868-MHz ISM spectrum, with a bandwidth of 125 kHz and a maximum transmit power of 14~dBm, which is the same as SigFox. 

At the \gls{phy} layer, it implements \gls{css} modulation which guarantees robustness to interference.
LoRa devices can choose different \glspl{sf}, with SF~$\in\{7,\dots,12\}$, which is a function of the number of bits sent per symbol.
Notably, the SF is inversely proportional to the raw data rate $R_s$ (up to around 6.5 kbps with SF7), i.e., 
\begin{equation}
    R_s=\text{SF}\cdot{B}/{2^{\text{SF}}},
\end{equation}
where $B$ is the bandwidth.
The SF also determines the transmission duration, i.e., the \gls{toa}, computed~as
\begin{equation}
    T = \frac{2^{\rm SF}}{B} \left( 8 + \max\left( 5 \, \left \lceil \frac{8\, L - 4\, \rm{SF} +24}{4\, \rm{SF}} \right \rceil ,0\right) \right),
    \label{eq:toa}
\end{equation}
where $L$ is the size of the message in bytes~\cite{9431912}. 
Transmissions with a higher $\rm{SF}$ require more time, which allows a reduced sensitivity at the receiver (from $-132$ dBm with SF7 to $-143$ dBm with SF12, as reported in~\cref{tab:params}.) and wider coverage  (up to 14 km with SF12). 
Generally, the SF is assigned based on the power level, where each device uses the lowest possible SF such that the received power is still above the gateway sensitivity. 
However, if multiple devices operate in similar conditions, they will select the same $\rm{SF}$, which increases the collision probability.
This is especially true in the NTN-IoT scenario, where devices tend to choose the highest possible SF to maximize the communication range, which may create interference.
Based on the assumption of quasi-orthogonality among different \glspl{sf}~\cite{chiani2019lora}, we propose a new method (referred to as LoRa+ in the rest of the paper) where end devices scramble across different SFs to reduce the impact of interference, regardless of the value of the sensitivity.

\subsection{NB-IoT}

\gls{nbiot} is an \gls{lpwan} technology designed, developed and standardized by the 3GPP.
\gls{nbiot} devices use \gls{ofdma} with 180 kHz of bandwidth, and a transmit power of 23 dBm. The subcarrier spacing is 15~kHz in downlink, and 15 or 3.75~kHz (15 kHz) for single-tone (multi-tone) transmissions in uplink.

NB-IoT supports repetition coding, with up to $2048$ (128) repetitions in downlink (uplink), which achieves coverage extension up to 10 km. Also, it improves the receiver sensitivity via coherent addition of the symbols and incoherent addition of thermal noise, but simultaneously increases the system latency. Notably, the sensitivity decreases by 2.8 dB whenever the number of repetitions is doubled~\cite{matz2020systematic}, as reported in~\cref{tab:params}. 
Unlike LoRa, NB-IoT is not immune to interference since repetition codes are not orthogonal, but allows for synchronization despite some additional cost and complexity in the~device. 

The channel access is based on Slotted ALOHA, which guarantees faster response time than other \gls{lpwan} technologies.
In this paper, with the assumptions of \gls{qpsk} modulation, code rate of $1/3$, and around $30\%$ of the uplink resources reserved, NB-IoT supports a data rate up to 90 kbps as considered in~\cite{malik2018radio}.

\section{System Model}
In this section we introduce our scenario (Sec.~\ref{sub:system-level}), and the link-level model including channel model, the signal detection policy and the traffic model (Sec.~\ref{sub:link-level}).


\subsection{Scenario}
\label{sub:system-level}

Our scenario consists of a ground-to-air/space uplink system in which \gls{leo} satellites (L), \glspl{hap} (H), \glspl{uav} (U), terrestrial gateways (TG), and \gls{iot} devices (ID) form a 3D network. Specifically: 
\begin{itemize}
    \item LEO satellites are deployed at $h=600$ km, and offer several advantages like huge coverage and good \gls{los} connectivity, at the expense of some delays due to the very long distance.
    \item HAPs are deployed in the stratosphere at $h=20$ km, and implement solar charging technology to provide long-life and stable wireless connectivity. 
    \item UAVs fly at $h=0.6$ km, and guarantee lower delay and installation/management costs than HAPs. However, they provide limited coverage, and incur significant energy consumption for propulsion and~hovering.
\end{itemize}
In this context, the availability of multi-layered networks can provide better coverage and flexibility compared to standalone deployments.
Based on our initial results in~\cite{wang2021potential}, in Sec.~\ref{sec:results} we will study the case of HAP relays for an upstream LEO satellite connected to the core network.
We assume that each NTN platform is equipped with multiple receivers working in parallel, where the center frequency of each receive path can be individually configured. Also, the mobility of NTN platforms is neglected.

IDs and TGs are uniformly distributed with a density $\rho_{\rm ID}$ and $\rho_{\rm TG}$, respectively, over an \gls{aoi} $A$, which is a circular area of radius $r$ split in cells of equal size.

\begin{table*}[t!] \label{tab:params1}
\renewcommand{\arraystretch}{1.7}
\caption{System parameters.}
    \label{tab:params1}
    \centering
\begin{tabular}{|l|ccccccccc|c|c|l|l|}
\hline
{LPWAN technology} & \multicolumn{4}{c|}{LoRa } & \multicolumn{3}{c|}{NB-IoT } & \multicolumn{3}{c|}{Sigfox} &\multicolumn{1}{c}{Relay}&\multicolumn{2}{|c|}{Other Parameters}\\ \cline{1-14}
Tx power ($P_t)$~[dBm] &  \multicolumn{4}{c|}{14 } & \multicolumn{3}{c|}{23 } & \multicolumn{3}{c|}{14} &\multicolumn{1}{c|}{52}&Altitude of UAV/HAP/LEO ($h$)~[km]  &  0.6/20/600\\ \cline{1-14}
Carrier frequency~[GHz] &  \multicolumn{4}{c|}{0.868} & \multicolumn{3}{c|}{0.900} & \multicolumn{3}{c|}{0.868} &\multicolumn{1}{c|}{38}&Additonal pathloss for (\gls{los},  NLOS)~[dB]   & (0.0154,18.4615)\\ \cline{1-14}
Bandwidth ($B$)~[MHz] &  \multicolumn{4}{c|}{0.125} & \multicolumn{3}{c|}{0.18} & \multicolumn{3}{c|}{0.2} &\multicolumn{1}{c|}{400}&Nakagami fading factor ($m_0$)  & 15\\ \cline{1-14}
Tx. antenna gain~[dB] &  \multicolumn{4}{c|}{2.15 } & \multicolumn{3}{c|}{0 } & \multicolumn{3}{c|}{2.15} &\multicolumn{1}{c|}{37.9} & Shadowed-Rician fading factor ($\omega, b_0 , m$)& (1.29, 0.158, 19.4)  \\ \cline{1-14}
Rx. antenna gain~[dB] &  \multicolumn{4}{c|}{8} & \multicolumn{3}{c|}{8 } & \multicolumn{3}{c|}{8} &\multicolumn{1}{c|}{0} & ID transmission rate $\lambda$ [tx/s] & 1/1800 \\ \cline{1-14}
Receiver noise figure ($\rm NF$)  [dB]&  \multicolumn{4}{c|}{3} & \multicolumn{3}{c|}{3} & \multicolumn{3}{|c|}{3} &\multicolumn{1}{c|}{0} & Max. payload size [byte] &12\\ \cline{1-14}


\end{tabular}
\end{table*}

\subsection{Link-Level Model}
\label{sub:link-level}
\paragraph{Channel model} 
\label{par:channel_model}
The received power $P_{ij}$ from transmitter $i$ to receiver $j$, $(i,j)\in\{\text{TG,U,H,L}\}$, is expressed~as
\begin{equation}
    P_{ij} =  P_{t_i} \,\mathrm{PL}_{ij} \,G_{ij}\, ||h_{ij}||^2,
    \label{eq:P}
\end{equation}
where 
$\mathrm{P}_{t_i}$ is the transmit power (which depends on the adopted \gls{lpwan} technology),
$G_{ij}$ is the cumulative antenna gain,
and $||h_{ij}||^2$ is the fading.
The path loss $\mathrm{PL}_{ij}$ depends on the type of link and, besides free-space path loss, accounts for atmospheric attenuation as described in~\cite{giordani2020satellite}.
In this work different channel models are used based on the link: 
(i) for the ground-to-ground (ID-TG) link we use the link performance model described in~\cite{magrin2017performance}, which computes the interference at reduced complexity via pairs of look-up tables; (ii) the ground-to-air (ID-\{U,H\}) link is modeled using a Nakagami-$m0$ fading model, as done in~\cite{lei2019safeguarding}; (iii) for the ground-to-space (ID-L) link the fading is based on a Shadowed-Rician model~\cite{loo1985}.
Channel parameters are listed in~\cref{tab:params1}. 
Then, the \gls{snr} $\gamma_{ij}$ is:
\begin{equation}
    \gamma_{ij} = P_{ij}/(BN_0+\text{NF}),
\end{equation}
where $B$ is the bandwidth, $N_0$ is the thermal noise power spectral density, and NF is the noise figure.
In case the \gls{hap} acts as a relay of an upstream LEO satellite in a multi-layered system~\cite{wang2021potential}, we implement a \gls{df} protocol where the SNR is constrained by the weakest link, i.e., 
\begin{equation}
	\gamma_{DF}=\min (\gamma_{ij}), \: (i,j)\in\{\text{ID,TG,U,H,L}\}.
	\label{eq:e2e-snr}
\end{equation} 

\paragraph{Signal detection policy} 
\label{par:signal_detection_policy}

Let $S$ be the receiver sensitivity as reported in~\cref{tab:params} for the different LPWAN technologies.
Successful packet transmission is subject to the following condition:
\begin{equation}
    P_{ij} \geq S.
    \label{eq:succ}
\end{equation}
While SigFox and NB-IoT are in a noise-limited regime, for LoRa the sensitivity depends on the SF. Therefore, a packet with SF$k$, $k\in\{7,\dots,12\}$, is correctly decoded if, for every set of interfering packets with the same SF, the received power is above the sensitivity threshold $S_k$~\cite{magrin2017performance}.



\paragraph{Traffic model}
We refer to the \gls{mar} model, introduced in~\cite{3gpp}. 
Hence, the payload size at the application is stochastic, and follows a Pareto distribution with 12 bytes of maximum size as per SigFox capacity limitations.
In addition, IDs transmit IoT data at constant periodicity, modeled as a Possion distribution of rate $\lambda= 1/1800$ transmissions/s.

\begin{figure*}[t]
    \centering
	\begin{subfigure}[b]{\linewidth}
		\centering
		\setlength\fwidth{\columnwidth}
%
%

\definecolor{chocolate2168224}{RGB}{216,82,24}
\definecolor{darkcyan0113188}{RGB}{0,113,188}
\definecolor{darkgray176}{RGB}{176,176,176}
\definecolor{goldenrod23617631}{RGB}{236,176,31}
\definecolor{purple12546141}{RGB}{125,46,141}

\begin{tikzpicture}
\pgfplotsset{every tick label/.append style={font=\scriptsize}}

\pgfplotsset{compat=1.11,
	/pgfplots/ybar legend/.style={
		/pgfplots/legend image code/.code={%
			\draw[##1,/tikz/.cd,yshift=-0.25em]
			(0cm,0cm) rectangle (10pt,0.6em);},
	},
}

\begin{axis}[%
width=0,
height=0,
at={(0,0)},
scale only axis,
xmin=0,
xmax=0,
xtick={},
ymin=0,
ymax=0,
ytick={},
axis background/.style={fill=white},
legend style={legend cell align=left,
              align=center,
              draw=white!15!black,
              at={(0, 0)},
              anchor=center,
              /tikz/every even column/.append style={column sep=1em}},
legend columns=6,
]

\addplot [semithick, blue, mark=triangle, mark size=2, mark options={solid}]
table[row sep=crcr]{%
	0	0\\
};
\addlegendentry{UAV-LoRa}

\addplot [semithick, blue, dashed, mark=pentagon, mark size=2, mark options={solid}, dashdotdotted]table[row sep=crcr]{%
	0	0\\
};
\addlegendentry{HAP-LoRa}

\addplot [semithick, blue, mark=square, mark size=2, mark options={solid}, dashed]
  table[row sep=crcr]{%
	0	0\\
};
\addlegendentry{LEO-LoRa}

\addplot [semithick, red, mark=triangle, mark size=2, mark options={solid}]
table[row sep=crcr]{%
	0	0\\
};
\addlegendentry{UAV-NB}

\addplot [semithick, red, dashed, mark=pentagon, mark size=2, mark options={solid},  dashdotdotted]
table[row sep=crcr]{%
	0	0\\
};
\addlegendentry{HAP-NB}

\addplot [semithick, red, mark=square, mark size=2, mark options={solid}, dashed]
table[row sep=crcr]{%
	-0	0\\
};
\addlegendentry{LEO-NB}

\addplot [semithick, green, mark=triangle, mark size=2, mark options={solid}]
table[row sep=crcr]{%
	0	0\\
};
\addlegendentry{UAV-LoRa+}

\addplot [semithick, green, dashed, mark=pentagon, mark size=2, mark options={solid},  dashdotdotted]
table[row sep=crcr]{%
	0	0\\
};
\addlegendentry{HAP-LoRa+}

\addplot [semithick, green, mark=square, mark size=2, mark options={solid}, dashed]
table[row sep=crcr]{%
	-0	0\\
};
\addlegendentry{LEO-LoRa+}

\addplot [semithick, black, mark=triangle, mark size=2, mark options={solid}]
table[row sep=crcr]{%
	0	0\\
};
\addlegendentry{UAV-SigFox}

\addplot [semithick, black, dashed, mark=pentagon, mark size=2, mark options={solid},  dashdotdotted]
table[row sep=crcr]{%
	0	0\\
};
\addlegendentry{HAP-SigFox}

\addplot [semithick, black, mark=square, mark size=2, mark options={solid}, dashed]
table[row sep=crcr]{%
	-0	0\\
};
\addlegendentry{LEO-SigFox}

,\end{axis}
\end{tikzpicture}%
	\end{subfigure}
    \vskip 0.5cm
    \centering 
	\subfloat[Goodput.]
	{
		\label{fig.iotcom}
\begin{tikzpicture}

\definecolor{darkgray176}{RGB}{176,176,176}
\definecolor{yellow}{RGB}{255,255,0}

\begin{axis}[
width = \textwidth/2.3,
height = 5cm,
log basis x={10},
log basis y={10},
tick align=outside,
tick pos=left,
x grid style={darkgray176},
xmajorgrids,
xmin=1, xmax=1000000,
xmode=log,
xtick style={color=black},
xlabel={Number of IDs},
y grid style={darkgray176},
ymin=10, ymax=2000000,
ymode=log,
ymajorgrids,
ytick style={color=black},
ylabel={Goodput [bytes/hour]},
]
\addplot [semithick, blue, mark=triangle, mark size=2, mark options={solid}]
table {%
1.25892541179417 24
1.99526231496888 24
3.16227766016838 48
5.01187233627272 72
7.94328234724282 96
12.5892541179417 168
19.9526231496888 240
31.6227766016838 384
50.1187233627272 600
79.4328234724281 936
125.892541179417 1464
199.526231496888 2304
316.227766016838 3648
501.187233627272 5736
794.328234724281 8928
1258.92541179417 14136
1995.26231496888 22392
3162.27766016838 35160
5011.87233627273 55608
7943.28234724281 84816
12589.2541179417 126624
19952.6231496888 188664
31622.7766016838 264288
50118.7233627273 350832
79432.8234724282 414384
125892.541179417 409910.4768
199526.231496888 305364.61824
316227.766016838 141278.76576
501187.233627272 30247.54368
794328.234724282 2012.70432
};
\addplot [semithick, blue, dashed, mark=pentagon, mark size=2, mark options={solid}, dashdotdotted]
table {%
1.25892541179417 24
1.99526231496888 24
3.16227766016838 48
5.01187233627272 72
7.94328234724282 96
12.5892541179417 168
19.9526231496888 240
31.6227766016838 384
50.1187233627272 600
79.4328234724281 936
125.892541179417 1464
199.526231496888 2256
316.227766016838 3648
501.187233627272 5784
794.328234724281 9120
1258.92541179417 14472
1995.26231496888 22560
3162.27766016838 35184
5011.87233627273 55728
7943.28234724281 85464
12589.2541179417 128232
19952.6231496888 190920
31622.7766016838 264552
50118.7233627273 352704
79432.8234724282 406728
125892.541179417 409025.98656
199526.231496888 289691.93088
316227.766016838 136580.4
501187.233627272 26380.01424
794328.234724282 1463.78496
};
\addplot [semithick, blue, mark=square, mark size=2, mark options={solid}, dashed]
table {%
1.25892541179417 24
1.99526231496888 24
3.16227766016838 48
5.01187233627272 72
7.94328234724282 96
12.5892541179417 168
19.9526231496888 240
31.6227766016838 384
50.1187233627272 600
79.4328234724281 936
125.892541179417 1464
199.526231496888 2304
316.227766016838 3648
501.187233627272 5784
794.328234724281 8976
1258.92541179417 14184
1995.26231496888 22272
3162.27766016838 35040
5011.87233627273 54696
7943.28234724281 82272
12589.2541179417 122088
19952.6231496888 176832
31622.7766016838 242040
50118.7233627273 300120
79432.8234724282 325656
125892.541179417 278266.42944
199526.231496888 168010.28928
316227.766016838 57327.34656
501187.233627272 7792.7832
794328.234724282 182.97312
};
\addplot [semithick, green, mark=triangle, mark size=2, mark options={solid}]
table {%
1.25892541179417 24
1.99526231496888 24
3.16227766016838 48
5.01187233627272 72
7.94328234724282 96
12.5892541179417 168
19.9526231496888 240
31.6227766016838 384
50.1187233627272 600
79.4328234724281 936
125.892541179417 1464
199.526231496888 2304
316.227766016838 3648
501.187233627272 5688
794.328234724281 9168
1258.92541179417 14424
1995.26231496888 22752
3162.27766016838 35736
5011.87233627273 56184
7943.28234724281 88680
12589.2541179417 135912
19952.6231496888 208680
31622.7766016838 315924
50118.7233627273 462432
79432.8234724282 644856
125892.541179417 829231.34976
199526.231496888 941280.4608
316227.766016838 868177.86528
501187.233627272 575049.67584
794328.234724282 223684.6392
};
\addplot [semithick, green, dashed, mark=pentagon, mark size=2, mark options={solid}, dashdotdotted]
table {%
1.25892541179417 24
1.99526231496888 24
3.16227766016838 48
5.01187233627272 72
7.94328234724282 96
12.5892541179417 168
19.9526231496888 240
31.6227766016838 384
50.1187233627272 600
79.4328234724281 936
125.892541179417 1464
199.526231496888 2304
316.227766016838 3648
501.187233627272 5784
794.328234724281 9072
1258.92541179417 14520
1995.26231496888 22944
3162.27766016838 36048
5011.87233627273 56544
7943.28234724281 88296
12589.2541179417 136920
19952.6231496888 211440
31622.7766016838 316584
50118.7233627273 464784
79432.8234724282 642816
125892.541179417 823735.9104
199526.231496888 936523.50144
316227.766016838 862823.9136
501187.233627272 564601.57392
794328.234724282 217829.49936
};
\addplot [semithick, green, mark=square, mark size=2, mark options={solid}, dashed]
table {%
1.25892541179417 24
1.99526231496888 24
3.16227766016838 48
5.01187233627272 72
7.94328234724282 96
12.5892541179417 168
19.9526231496888 240
31.6227766016838 384
50.1187233627272 600
79.4328234724281 936
125.892541179417 1416
199.526231496888 2304
316.227766016838 3600
501.187233627272 5784
794.328234724281 9120
1258.92541179417 14304
1995.26231496888 22608
3162.27766016838 35712
5011.87233627273 55848
7943.28234724281 87000
12589.2541179417 135240
19952.6231496888 204192
31622.7766016838 302376
50118.7233627273 428472
79432.8234724282 574536
125892.541179417 689568.89088
199526.231496888 720116.32128
316227.766016838 572399.35104
501187.233627272 306920.20944
794328.234724282 87186.69168
};
\addplot [semithick, red, mark=triangle, mark size=2, mark options={solid}]
table {%
1.25892541179417 24
1.99526231496888 24
3.16227766016838 48
5.01187233627272 72
7.94328234724282 96
12.5892541179417 168
19.9526231496888 240
31.6227766016838 384
50.1187233627272 600
79.4328234724281 936
125.892541179417 1464
199.526231496888 2304
316.227766016838 3648
501.187233627272 5784
794.328234724281 9120
1258.92541179417 14472
1995.26231496888 22896
3162.27766016838 36000
5011.87233627273 56928
7943.28234724281 89448
12589.2541179417 140904
19952.6231496888 218784
31622.7766016838 335280
50118.7233627273 505368
79432.8234724282 740112
125892.541179417 1040073.5232
199526.231496888 1344863.65056
316227.766016838 1568052.25632
501187.233627272 1511338.14624
794328.234724282 1091892.0936
};
\addplot [semithick, red, dashed, mark=pentagon, mark size=2, mark options={solid},  dashdotdotted]
table {%
1.25892541179417 24
1.99526231496888 24
3.16227766016838 48
5.01187233627272 72
7.94328234724282 96
12.5892541179417 168
19.9526231496888 240
31.6227766016838 384
50.1187233627272 600
79.4328234724281 936
125.892541179417 1464
199.526231496888 2304
316.227766016838 3648
501.187233627272 5784
794.328234724281 9072
1258.92541179417 14520
1995.26231496888 22656
3162.27766016838 35904
5011.87233627273 56928
7943.28234724281 89472
12589.2541179417 140328
19952.6231496888 217464
31622.7766016838 334296
50118.7233627273 505752
79432.8234724282 736872
125892.541179417 1038217.54368
199526.231496888 1348931.19552
316227.766016838 1567724.46336
501187.233627272 1502044.53072
794328.234724282 1102960.52064
};
\addplot [semithick, red, mark=square, mark size=2, mark options={solid}, dashed]
table {%
1.25892541179417 24
1.99526231496888 24
3.16227766016838 48
5.01187233627272 72
7.94328234724282 96
12.5892541179417 168
19.9526231496888 240
31.6227766016838 384
50.1187233627272 600
79.4328234724281 936
125.892541179417 1464
199.526231496888 2256
316.227766016838 3600
501.187233627272 5688
794.328234724281 8736
1258.92541179417 13152
1995.26231496888 20328
3162.27766016838 30600
5011.87233627273 42096
7943.28234724281 57624
12589.2541179417 65760
19952.6231496888 69672
31622.7766016838 54408
50118.7233627273 25680
79432.8234724282 6012
125892.541179417 492.99456
199526.231496888 22.98048
316227.766016838 0
501187.233627272 0
794328.234724282 0
};
\addplot [semithick, black, mark=triangle, mark size=2, mark options={solid}]
table {%
1.25892541179417 24
1.99526231496888 24
3.16227766016838 48
5.01187233627272 72
7.94328234724282 96
12.5892541179417 168
19.9526231496888 240
31.6227766016838 384
50.1187233627272 600
79.4328234724281 936
125.892541179417 1464
199.526231496888 2304
316.227766016838 3648
501.187233627272 5784
794.328234724281 9168
1258.92541179417 14520
1995.26231496888 22992
3162.27766016838 36192
5011.87233627273 54720
7943.28234724281 44412
12589.2541179417 3528
19952.6231496888 0
31622.7766016838 0
50118.7233627273 0
79432.8234724282 0
125892.541179417 0
199526.231496888 0
316227.766016838 0
501187.233627272 0
794328.234724282 0
};
\addplot [semithick, black, dashed, mark=pentagon, mark size=2, mark options={solid},  dashdotdotted]
table {%
1.25892541179417 24
1.99526231496888 24
3.16227766016838 48
5.01187233627272 72
7.94328234724282 96
12.5892541179417 168
19.9526231496888 240
31.6227766016838 384
50.1187233627272 600
79.4328234724281 936
125.892541179417 1464
199.526231496888 2304
316.227766016838 3648
501.187233627272 5784
794.328234724281 9168
1258.92541179417 14520
1995.26231496888 22992
3162.27766016838 36432
5011.87233627273 53976
7943.28234724281 46752
12589.2541179417 1968
19952.6231496888 0
31622.7766016838 0
50118.7233627273 0
79432.8234724282 0
125892.541179417 0
199526.231496888 0
316227.766016838 0
501187.233627272 0
794328.234724282 0
};
\addplot [semithick, black, mark=square, mark size=2, mark options={solid}, dashed]
table {%
1.25892541179417 24
1.99526231496888 24
3.16227766016838 48
5.01187233627272 72
7.94328234724282 96
12.5892541179417 168
19.9526231496888 240
31.6227766016838 384
50.1187233627272 600
79.4328234724281 936
125.892541179417 1464
199.526231496888 2304
316.227766016838 3648
501.187233627272 5784
794.328234724281 9168
1258.92541179417 14520
1995.26231496888 22992
3162.27766016838 36432
5011.87233627273 54744
7943.28234724281 40632
12589.2541179417 1392
19952.6231496888 0
31622.7766016838 0
50118.7233627273 0
79432.8234724282 0
125892.541179417 0
199526.231496888 0
316227.766016838 0
501187.233627272 0
794328.234724282 0
};
\end{axis}

\end{tikzpicture}
	}
     \subfloat[Average success probability.]
     {
        \label{fig:iotcomp}
        \input{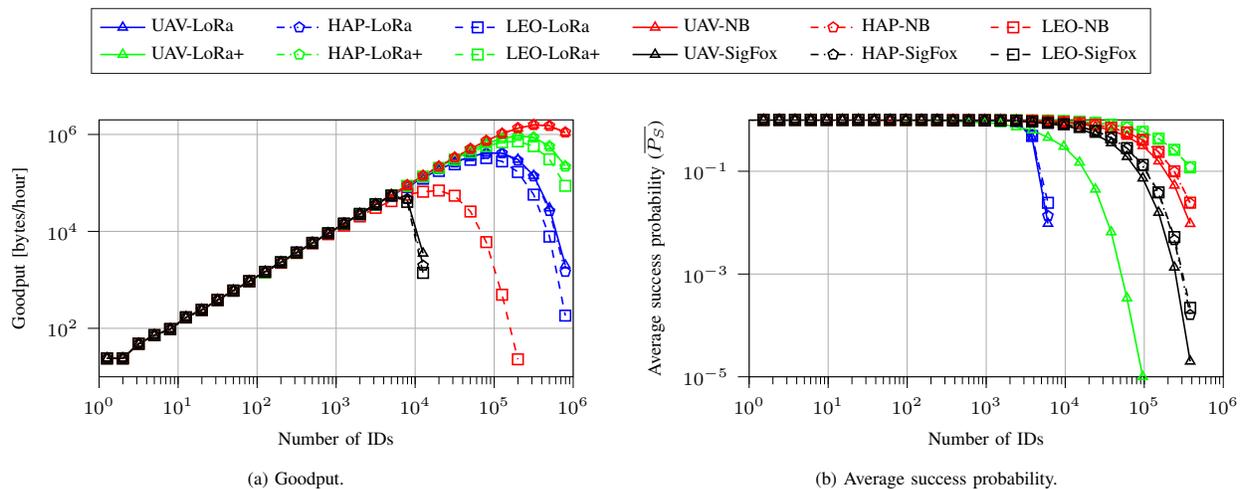}    
     }
\caption{Network capacity and average success probability of different \gls{lpwan} technologies, vs. the number of IDs, considering an \gls{aoi} of radius $r=0.35$~km.}
\label{fig:capacity}
\end{figure*}

\section{Optimized Offloading}
\label{sec_offloading}
Besides (inter)connecting IDs, NTNs can act as complementary computing servers for processing IoT data, in addition to (or in place of) TGs in hot-spot (or rural) areas, respectively~\cite{nguyen20216g}.
As a case study we focus on LoRa, and consider the scenario in which IDs offload data to a LEO satellite with probability $\eta$, while with probability $(1-\eta)$ the data is processed onboard the TG they are connected to~\cite{traspadini2022uavhapassisted}. 
We introduce the following assumptions: 
\begin{enumerate}
     \item For ground-to-ground (ID-TG) communication, IDs use SF$k$, $k\in\{7,\dots,12\}$, based on the model in~\cref{sub:LoRa}.
     \item For ground-to-space (ID-L) communication, IDs use SF$v$, $v\in\{\text{SF}_{\rm min},\dots,12\}$, where SF$_{\rm min}=\{7,9,11\}$ is proportional to the quality of the ID-L link. This approach prevents IDs from choosing the same SF in the attempt to maximize the coverage range towards the LEO satellite.
 \end{enumerate} 
For a given SF$k$ in the ID-TG link, the optimal offloading factor $\eta_k^*$ must be dimensioned to maximize the success probability $P_{S_k}$, i.e., the probability that there are no collisions (or there are no IDs using the same SF) in the \gls{toa}. We have that
\begin{equation}
    P_{S_k}(\eta_k) = \left[(1-\eta_k) {P_{S_k}}^{\rm TG} + \eta_k {P_{S}}^{\rm L}\right].
    \label{eq:P_S_k}
\end{equation}
In Eq.~\eqref{eq:P_S_k}, ${P_{S}}^{\rm L}$ is the success probability in the ID-L link, and ${P_{S_k}}^{\rm TG}$ is the success probability in the ID-TG link,~i.e., 
\begin{equation}
    {P_{S_k}}^{\rm TG} (\eta_k) = e^{-(1-\eta_k)T_{k} \lambda |\mathfrak{D}_{k} |},
\end{equation}
where $|\mathfrak{D}_{k} |$ is the number of devices that use SF$k$ towards the TG, $T_{k}$ is the \gls{toa} using SF$k$ (see Eq.~\eqref{eq:toa}), and $\lambda$ is the rate at which IDs generate data.
Then, the optimization problem~is:
\begin{subequations}
  \label{eq:opt_problem_2}
  \begin{alignat}{2}
      &\argmax_{{\eta_k}} &\quad&P_{S_k}(\eta_k), \\
      &\text{subject to} & & \eta_k \in [0,1].
  \end{alignat}
\end{subequations}

The problem in~\eqref{eq:opt_problem_2} is subject to the optimization of ${P_{S}}^{\rm L}$, which requires that the IDs offloading data to the LEO satellite (with probability $\eta^*$) choose their SFs so as to maximize the success probability in the ID-L link. This is formalized as:
\begin{subequations}
	\label{eq:opt_problem_1}
	\begin{alignat}{2}
		 {P_{S}}^{\rm L}(\eta_k) = & \argmax_{\alpha_v} && \sum\nolimits_{v = \rm{SF}_{min}}^{12} \alpha_v \cdot e^{-\alpha_v T_{v} \lambda |\Delta(\eta_k)| } , \\
		&\text{subject to} & & \sum\nolimits_{v = \rm{SF}_{min}}^{12} \alpha_v =1, \label{eq:unity}\\
        &  & & \: |\Delta(\eta_k)| = \sum_{\substack{j=7 \\ j\neq k}}^{12} \eta_j^* |\mathfrak{D}_j|+\eta_k |\mathfrak{D}_k|, \label{eq:totdevs}\\
		&  & & \: \alpha_v \in [0,1],
	\end{alignat}
\end{subequations}
 where $\alpha_v$ is the probability that an ID chooses SF$v$ in the ID-L link, and $|\Delta(\eta_k)|$ denotes the total number of IDs that offload data to the LEO satellite. 
 The results of the optimization problem in \eqref{eq:opt_problem_2} will be presented in~\cref{sub:res-opt}.

\section{Performance Evaluation}
\label{sec:results}
In this section, we compare the performance of the \gls{lpwan} technologies introduced in Sec.~\ref{sec:lpwan} for different \gls{ntn} configurations, in terms of network capacity (\cref{sub:capacity}), success probability (\cref{sub:success_probability}), and coverage (\cref{sub:coverage}).
Then, in \cref{sub:res-opt} we validate the offloading framework described in~\cref{sec_offloading} based on LoRa.

\subsection{Network Capacity}
\label{sub:capacity}
We consider a scenario with up to $10^6$ IDs uniformly distributed in an \gls{aoi} of radius $r=0.35$~km as defined by the coverage range of the UAV, i.e., the most constrained \gls{ntn} platform.
In \cref{fig:capacity} we see that, when the number of IDs is lower than $10^3$, the interference is negligible and all \gls{lpwan} technologies guarantee similar values of goodput, with high success probability.
For SigFox, as the ID density increases, and despite using \gls{fhss} to reduce the interference, the goodput is eventually constrained by the limited capacity available at the PHY layer (below 100 bps on average), and is up to 10 times lower than its competitors.
On the other hand, NB-IoT provides the highest goodput for \gls{uav}- and \gls{hap}-enabled networks (up to $1.5\cdot10^6$~bytes/hour) given the higher data rate at the PHY-layer (up to 90 kbps with QPSK modulation). 
However, the goodput drops below $10^5$~bytes/hour when LEO satellite links are considered, where LoRa shows instead superior performance, with a goodput of $3\cdot 10^5$~bytes/hour.
In fact, the flexibility of LoRa allows IDs to select higher SFs to operate at much lower sensitivity compared to NB-IoT (the gap is up to 20 dB), thus increasing the communication range.

In addition, we evaluate the performance of LoRa+ in which SFs are assigned based on the model described in~\cref{sub:LoRa} to minimize interference. This approach increases the capacity by about 50\% compared to the baseline LoRa implementation, and the maximum goodput is close to $10^6$~bytes/hour.

\subsection{Success Probability} 
\label{sub:success_probability}

\begin{figure}[t!]
    \centering
    \begin{subfigure}[b]{\linewidth}
        \centering
       \setlength\fwidth{\columnwidth}
%
%

\definecolor{chocolate2168224}{RGB}{216,82,24}
\definecolor{darkcyan0113188}{RGB}{0,113,188}
\definecolor{darkgray176}{RGB}{176,176,176}
\definecolor{goldenrod23617631}{RGB}{236,176,31}
\definecolor{purple12546141}{RGB}{125,46,141}

\begin{tikzpicture}
\pgfplotsset{every tick label/.append style={font=\scriptsize}}

\pgfplotsset{compat=1.11,
	/pgfplots/ybar legend/.style={
		/pgfplots/legend image code/.code={%
			\draw[##1,/tikz/.cd,yshift=-0.25em]
			(0cm,0cm) rectangle (10pt,0.6em);},
	},
}

\begin{axis}[%
width=0,
height=0,
at={(0,0)},
scale only axis,
xmin=0,
xmax=0,
xtick={},
ymin=0,
ymax=0,
ytick={},
axis background/.style={fill=white},
legend style={legend cell align=left,
              align=center,
              draw=white!15!black,
              at={(0, 0)},
              anchor=center,
              /tikz/every even column/.append style={column sep=0.3em}},
legend columns=5,
]

\addplot [thick, darkcyan0113188, mark=diamond, mark size=2, mark options={solid}]
  table[row sep=crcr]{%
	0	0\\
};
\addlegendentry{Standalone ID-TG}


\addplot [thick, purple12546141, mark=square, mark size=2, mark options={solid}, dashed]
table[row sep=crcr]{%
	-0	0\\
};
\addlegendentry{Standalone ID-L}

\addplot [thick, goldenrod23617631, mark=pentagon*, mark size=2, mark options={solid}, dashed]
table[row sep=crcr]{%
	0	0\\
};
\addlegendentry{Multi-layered ID-H-L}

\end{axis}
\end{tikzpicture}%
    \end{subfigure}
    \centering 
    \subfloat{
\begin{tikzpicture}

\definecolor{chocolate2168224}{RGB}{216,82,24}
\definecolor{darkcyan0113188}{RGB}{0,113,188}
\definecolor{darkgray176}{RGB}{176,176,176}
\definecolor{goldenrod23617631}{RGB}{236,176,31}
\definecolor{purple12546141}{RGB}{125,46,141}

\begin{axis}[
width = \textwidth/2.3,
height = 5cm,
tick align=outside,
tick pos=left,
x grid style={darkgray176},
xlabel={Radius of the AoI ($r$)~[km]},
xmajorgrids,
xmin=0, xmax=20,
xtick style={color=black},
y grid style={darkgray176},
ylabel={Average success probability ($\overline{P_S}$)},
ymajorgrids,
ymin=0.8, ymax=1,
ytick style={color=black}
]
\addplot [semithick, darkcyan0113188, mark=diamond, mark size=2, mark options={solid}]
table {%
1 0.979864
2 0.981016
3 0.98378
4 0.983052
5 0.98488
6 0.985124
7 0.985332
8 0.98594
9 0.985692
10 0.986564
11 0.986268
12 0.986452
13 0.986804
14 0.986908
15 0.98638
16 0.98684
17 0.987044
18 0.98716
19 0.987152
20 0.987024
};
\addplot [semithick, chocolate2168224, mark=triangle*, mark size=2, mark options={solid}, dashed]
table {%
1 0
2 0
3 0
4 0
5 0
6 0
7 0
8 0
9 0
10 0
11 0
12 0
13 0
14 0
15 0
16 0
17 0
18 0
19 0
20 0
};
\addplot [semithick, goldenrod23617631, mark=pentagon*, mark size=2, mark options={solid}, dashed]
table {%
1 0.996952
2 0.996324
3 0.99494
4 0.993852
5 0.991872
6 0.98958
7 0.98672
8 0.982892
9 0.978692
10 0.972176
11 0.96524
12 0.957356
13 0.94746
14 0.93614
15 0.924072
16 0.909544
17 0.895912
18 0.88016
19 0.862972
20 0.845532
};
\addplot [semithick, purple12546141, mark=square, mark size=2, mark options={solid}, dashed]
table {%
1 0.940284
2 0.941468
3 0.940564
4 0.941272
5 0.94114
6 0.941476
7 0.940944
8 0.940968
9 0.940624
10 0.94068
11 0.940516
12 0.941272
13 0.940916
14 0.940344
15 0.940516
16 0.941004
17 0.941164
18 0.940388
19 0.94044
20 0.940676
};
\end{axis}

\end{tikzpicture}
    }
\caption{Average success probability vs. $r$, based on LoRa. We consider ground-to-ground (ID-TG) transmissions with the TG, or ground-to-space (ID-L) transmissions with the LEO satellite, possibly relayed via a HAP (ID-H-L).}
\label{fig.1gs}
\end{figure}
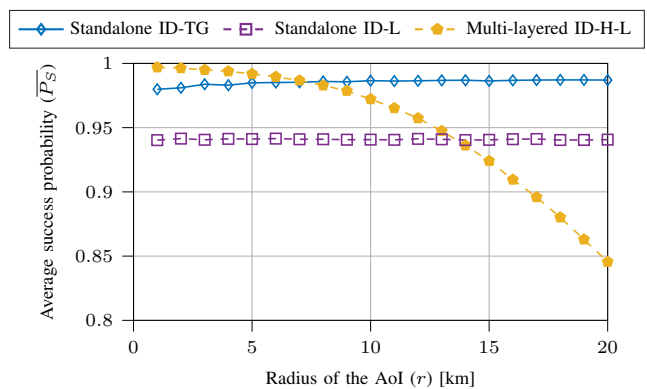

Similar trends can be observed in~\cref{fig:iotcomp}, which shows the average success probability based on the definition in Eq.~\eqref{eq:P_S_k}.

In addition, in~\cref{fig.1gs} we focus on LoRa, and consider a multi-layered network in which a HAP acts as a relay of an upstream LEO satellite (ID-H-L) vs. two standalone configurations in which IDs communicate with the TG (ID-TG) or the LEO satellite (ID-L).
In this scenario $\rho_{\rm TG}=1$~TG/km$^2$ and $\rho_{\rm ID} = 10$~ID/km$^2$.
We see that both TGs and LEO satellites can serve most ID requests with success, even though ID-TG involves more expensive network densification as the radius of the AoI increases.
Moreover, ID-H-L outperforms standalone ID-L by around 10\% for $r<15$~km, after which the scenario is constrained by the limited coverage area of the HAP, as explained in the next subsection.


\subsection{Network Coverage}
\label{sub:coverage}

In this third set of results we focus on the coverage performance of LoRa and NB-IoT, since SigFox was observed to provide insufficient capacity to support NTN-IoT.

In~\cref{tab:distance} we report the maximum achievable range $r$ and the minimum possible elevation angle $\theta$ for which the received power is higher than the lowest sensitivity.
We observe that LoRa outperforms NB-IoT under both metrics thanks to the lower sensitivity.
Moreover, LEO satellites provide the largest coverage area (up to $1\,450$ km), as they operate in \gls{los}~and suffer from less severe visibility constraints than other NTN platforms.
Interestingly, \glspl{uav} provide limited coverage compared to TGs. In fact, UAVs fly at low altitude, which implies that the elevation angle is very low: this means that the link is longer, which makes the signal experience more~attenuation.

\begin{figure}[t!]
    \centering
    \vspace{0.5cm}
    \begin{subfigure}[b]{\linewidth}
        \centering
        \setlength\fwidth{\columnwidth}
%
%

\definecolor{chocolate2168224}{RGB}{216,82,24}
\definecolor{darkcyan0113188}{RGB}{0,113,188}
\definecolor{darkgray176}{RGB}{176,176,176}
\definecolor{goldenrod23617631}{RGB}{236,176,31}
\definecolor{purple12546141}{RGB}{125,46,141}

\begin{tikzpicture}
\pgfplotsset{every tick label/.append style={font=\scriptsize}}

\pgfplotsset{compat=1.11,
	/pgfplots/ybar legend/.style={
		/pgfplots/legend image code/.code={%
			\draw[##1,/tikz/.cd,yshift=-0.25em]
			(0cm,0cm) rectangle (10pt,0.6em);},
	},
}

\begin{axis}[%
width=0,
height=0,
at={(0,0)},
scale only axis,
xmin=0,
xmax=0,
xtick={},
ymin=0,
ymax=0,
ytick={},
axis background/.style={fill=white},
legend style={legend cell align=left,
              align=center,
              draw=white!15!black,
              at={(0, 0)},
              anchor=center,
              /tikz/every even column/.append style={column sep=1em}},
legend columns=4,
]
\addplot [semithick, blue, mark=diamond, mark size=2, mark options={solid}, dashdotted]
table[row sep=crcr]{%
	0	0\\
};
\addlegendentry{GS-LoRa}

\addplot [semithick, blue, mark=triangle, mark size=2, mark options={solid}]
table[row sep=crcr]{%
	0	0\\
};
\addlegendentry{UAV-LoRa}

\addplot [semithick, blue, dashed, mark=pentagon, mark size=2, mark options={solid}, dashdotdotted]table[row sep=crcr]{%
	0	0\\
};
\addlegendentry{HAP-LoRa}

\addplot [semithick, blue, mark=square, mark size=2, mark options={solid}, dashed]
  table[row sep=crcr]{%
	0	0\\
};
\addlegendentry{LEO-LoRa}

\addplot [semithick, red, mark=diamond, mark size=2, mark options={solid}, dashdotted]
table[row sep=crcr]{%
	0	0\\
};
\addlegendentry{GS-NB}

\addplot [semithick, red, mark=triangle, mark size=2, mark options={solid}]
table[row sep=crcr]{%
	0	0\\
};
\addlegendentry{UAV-NB}

\addplot [semithick, red, dashed, mark=pentagon, mark size=2, mark options={solid},  dashdotdotted]
table[row sep=crcr]{%
	0	0\\
};
\addlegendentry{HAP-NB}

\addplot [semithick, red, mark=square, mark size=2, mark options={solid}, dashed]
table[row sep=crcr]{%
	-0	0\\
};
\addlegendentry{LEO-NB}

,\end{axis}
\end{tikzpicture}%
    \end{subfigure}
    \vskip 0.2cm
    \centering 
    \subfloat{
\begin{tikzpicture}

\definecolor{darkgray176}{RGB}{176,176,176}

\begin{axis}[
width = \textwidth/2.3,
height = 6cm,
tick align=outside,
tick pos=left,
x grid style={darkgray176},
xmin=0, xmax=1400,
xtick style={color=black},
xmajorgrids,
xlabel={Radius of the AoI ($r$)~[km]},
y grid style={darkgray176},
ymajorgrids,
ymin=0, ymax=200,
ytick style={color=black},
ylabel={Min. number of NTN platforms}
]
\addplot [semithick, blue, mark=diamond, mark size=2, mark options={solid}, dashdotted]
table {%
1.25892541179417 1
1.58489319246111 1
1.99526231496888 1
2.51188643150958 1
3.16227766016838 1
3.98107170553497 1
5.01187233627272 1
6.30957344480193 1
7.94328234724282 1
10 1
12.5892541179417 1
15.8489319246111 2
19.9526231496888 2
25.1188643150958 2
31.6227766016838 3
39.8107170553497 3
50.1187233627272 4
63.0957344480193 5
79.4328234724281 6
100 7
125.892541179417 9
158.489319246111 12
199.526231496888 14
251.188643150958 18
316.227766016838 23
398.107170553497 28
501.187233627272 36
630.957344480193 45
794.328234724281 56
1000 70
1258.92541179417 89
};
\addplot [semithick, blue, mark=triangle, mark size=2, mark options={solid}]
table {%
1.25892541179417 1
1.58489319246111 1
1.99526231496888 1
2.51188643150958 1
3.16227766016838 1
3.98107170553497 1
5.01187233627272 1
6.30957344480193 1
7.94328234724282 1
10 2
12.5892541179417 2
15.8489319246111 2
19.9526231496888 3
25.1188643150958 3
31.6227766016838 4
39.8107170553497 5
50.1187233627272 6
63.0957344480193 8
79.4328234724281 10
100 12
125.892541179417 15
158.489319246111 19
199.526231496888 24
251.188643150958 30
316.227766016838 38
398.107170553497 48
501.187233627272 60
630.957344480193 76
794.328234724281 95
1000 120
1258.92541179417 150
};
\addplot [semithick, blue, dashed, mark=pentagon, mark size=2, mark options={solid}, dashdotdotted]
table {%
1.25892541179417 1
1.58489319246111 1
1.99526231496888 1
2.51188643150958 1
3.16227766016838 1
3.98107170553497 1
5.01187233627272 1
6.30957344480193 1
7.94328234724282 1
10 1
12.5892541179417 1
15.8489319246111 1
19.9526231496888 1
25.1188643150958 1
31.6227766016838 1
39.8107170553497 1
50.1187233627272 1
63.0957344480193 1
79.4328234724281 1
100 1
125.892541179417 2
158.489319246111 2
199.526231496888 2
251.188643150958 3
316.227766016838 4
398.107170553497 4
501.187233627272 5
630.957344480193 7
794.328234724281 8
1000 10
1258.92541179417 13
};
\addplot [semithick, blue, mark=square, mark size=2, mark options={solid}, dashed]
table {%
1.25892541179417 1
1.58489319246111 1
1.99526231496888 1
2.51188643150958 1
3.16227766016838 1
3.98107170553497 1
5.01187233627272 1
6.30957344480193 1
7.94328234724282 1
10 1
12.5892541179417 1
15.8489319246111 1
19.9526231496888 1
25.1188643150958 1
31.6227766016838 1
39.8107170553497 1
50.1187233627272 1
63.0957344480193 1
79.4328234724281 1
100 1
125.892541179417 1
158.489319246111 1
199.526231496888 1
251.188643150958 1
316.227766016838 1
398.107170553497 1
501.187233627272 1
630.957344480193 1
794.328234724281 1
1000 1
1258.92541179417 1
};
\addplot [semithick, red, mark=diamond, mark size=2, mark options={solid}, dashdotted]
table {%
1.25892541179417 1
1.58489319246111 1
1.99526231496888 1
2.51188643150958 1
3.16227766016838 1
3.98107170553497 1
5.01187233627272 1
6.30957344480193 1
7.94328234724282 1
10 2
12.5892541179417 2
15.8489319246111 2
19.9526231496888 3
25.1188643150958 3
31.6227766016838 4
39.8107170553497 5
50.1187233627272 6
63.0957344480193 8
79.4328234724281 10
100 12
125.892541179417 15
158.489319246111 19
199.526231496888 23
251.188643150958 29
316.227766016838 37
398.107170553497 46
501.187233627272 58
630.957344480193 73
794.328234724281 92
1000 115
1258.92541179417 145
};
\addplot [semithick, red, mark=triangle, mark size=2, mark options={solid}]
table {%
1.25892541179417 1
1.58489319246111 1
1.99526231496888 1
2.51188643150958 1
3.16227766016838 1
3.98107170553497 1
5.01187233627272 1
6.30957344480193 1
7.94328234724282 2
10 2
12.5892541179417 2
15.8489319246111 3
19.9526231496888 3
25.1188643150958 4
31.6227766016838 5
39.8107170553497 6
50.1187233627272 8
63.0957344480193 10
79.4328234724281 12
100 15
125.892541179417 19
158.489319246111 24
199.526231496888 30
251.188643150958 37
316.227766016838 47
398.107170553497 59
501.187233627272 74
630.957344480193 93
794.328234724281 117
1000 148
1258.92541179417 186
};
\addplot [semithick, red, dashed, mark=pentagon, mark size=2, mark options={solid},  dashdotdotted]
table {%
1.25892541179417 1
1.58489319246111 1
1.99526231496888 1
2.51188643150958 1
3.16227766016838 1
3.98107170553497 1
5.01187233627272 1
6.30957344480193 1
7.94328234724282 1
10 1
12.5892541179417 1
15.8489319246111 1
19.9526231496888 1
25.1188643150958 1
31.6227766016838 1
39.8107170553497 1
50.1187233627272 1
63.0957344480193 1
79.4328234724281 1
100 2
125.892541179417 2
158.489319246111 2
199.526231496888 3
251.188643150958 3
316.227766016838 4
398.107170553497 5
501.187233627272 6
630.957344480193 7
794.328234724281 9
1000 12
1258.92541179417 14
};
\addplot [semithick, red, mark=square, mark size=2, mark options={solid}, dashed]
table {%
1.25892541179417 1
1.58489319246111 1
1.99526231496888 1
2.51188643150958 1
3.16227766016838 1
3.98107170553497 1
5.01187233627272 1
6.30957344480193 1
7.94328234724282 1
10 1
12.5892541179417 1
15.8489319246111 1
19.9526231496888 1
25.1188643150958 1
31.6227766016838 1
39.8107170553497 1
50.1187233627272 1
63.0957344480193 1
79.4328234724281 1
100 1
125.892541179417 1
158.489319246111 1
199.526231496888 1
251.188643150958 1
316.227766016838 1
398.107170553497 1
501.187233627272 2
630.957344480193 2
794.328234724281 2
1000 3
1258.92541179417 3
};
\end{axis}

\end{tikzpicture}
    }
\caption{Minimum number of platforms needed to cover the whole area vs. the radius of the service area.}
\label{fig.coverage}
\end{figure}
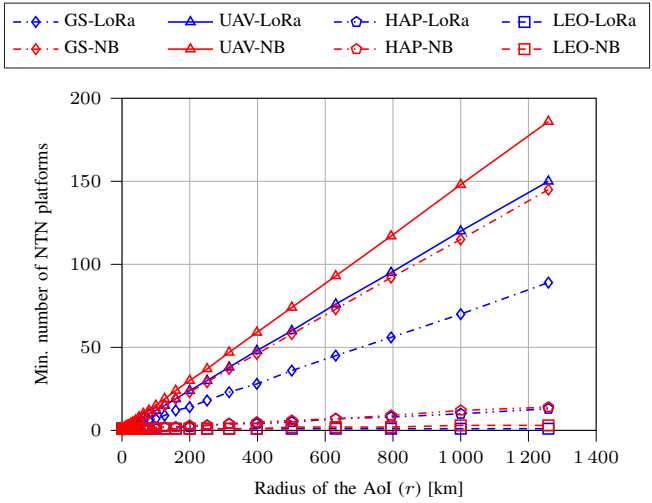

Similarly, \cref{fig.coverage} represents the minimum number of NTN platforms that need to be deployed to cover an AoI of radius $r$ while ensuring successful packet transmission as described in Eq.~\eqref{eq:succ}.
As expected, platforms at higher altitude like LEO satellites provide better coverage despite the resulting lower capacity as shown in~\cref{fig:capacity}, and NB-IoT needs more platforms to connect IDs compared to LoRa.


\begin{table}[b!] \label{tab:distance}
\renewcommand{\arraystretch}{1.7}
\caption{The maximum distance between devices and platforms}
    \label{tab:distance}
    \centering
\begin{tabular}{|c|c|c|c|c|}
\hline
\multirow{2}{*}{Platform}  &  \multicolumn{2}{c|}{LoRa} & \multicolumn{2}{c|}{NB-IoT}  \\ \cline{2-5}
 & $r$ [km] & $\theta$ [deg] & $r$ [km] & $\theta$ [deg]\\ \cline{1-5}
{TG} & {14.3} & N/A & {8.7} & N/A\\ \cline{1-5}
{UAV} & {8.4} & 4 & {6.8} & 5 \\ \cline{1-5}
{HAP} & {104.6} & 10.3 & {90.4} & 12 \\ \cline{1-5}
{LEO} & {1463.9} & 14.7 & {1278.8} & 48\\ \cline{1-5}
\end{tabular}
\end{table}

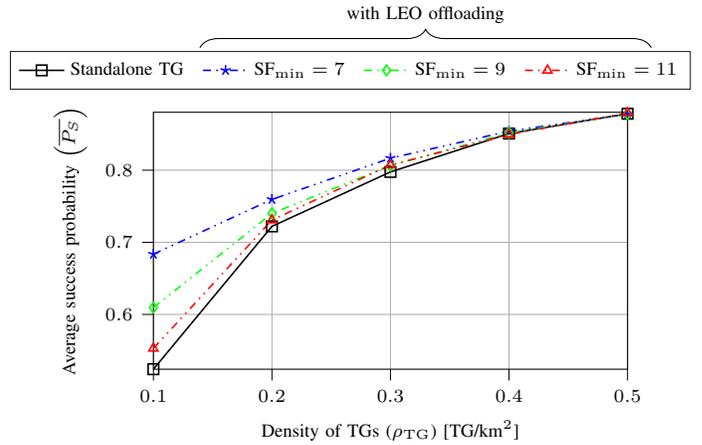
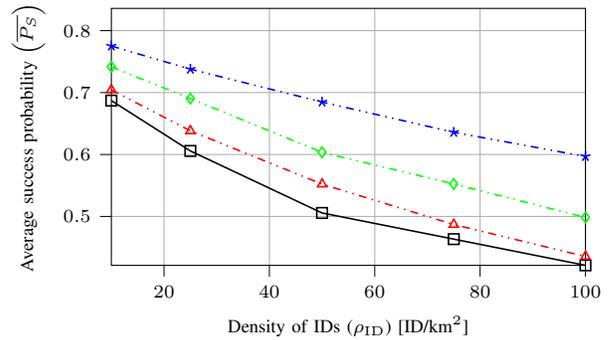
\begin{figure}[t!]
    \centering
    \begin{subfigure}[b]{\columnwidth}
        \centering
        \setlength\fwidth{\columnwidth}
%
%

\definecolor{chocolate2168224}{RGB}{216,82,24}
\definecolor{darkcyan0113188}{RGB}{0,113,188}
\definecolor{darkgray176}{RGB}{176,176,176}
\definecolor{goldenrod23617631}{RGB}{236,176,31}
\definecolor{purple12546141}{RGB}{125,46,141}

\begin{tikzpicture}

\draw [decorate,decoration={brace,mirror,amplitude=7pt},xshift=0pt,yshift=-0.2cm](4,0.5) -- (-2,0.5) node[black,midway,above,xshift=0cm,yshift=0.2cm] 
{\scriptsize with LEO offloading};

\pgfplotsset{every tick label/.append style={font=\scriptsize}}

\pgfplotsset{compat=1.11,
	/pgfplots/ybar legend/.style={
		/pgfplots/legend image code/.code={%
			\draw[##1,/tikz/.cd,yshift=-0.25em]
			(0cm,0cm) rectangle (10pt,0.6em);},
	},
}

\begin{axis}[%
width=0,
height=0,
at={(0,0)},
scale only axis,
xmin=0,
xmax=0,
xtick={},
ymin=0,
ymax=0,
ytick={},
axis background/.style={fill=white},
legend style={legend cell align=left,
              align=center,
              draw=white!15!black,
              at={(0, 0)},
              anchor=center,
              /tikz/every even column/.append style={column sep=0.5em}},
legend columns=6,
]

\addplot [semithick, black, mark=square, mark size=2, mark options={solid}]
table[row sep=crcr]{%
	0	0\\
};
\addlegendentry{Standalone TG}

\addplot [semithick, blue, dashed, mark=star, mark size=2, mark options={solid}, dashdotdotted]
table[row sep=crcr]{%
	0	0\\
};
\addlegendentry{SF$_{\rm min}=7$}

\addplot [semithick, green, mark=diamond, mark size=2, mark options={solid}, dashdotdotted]
table[row sep=crcr]{%
	0	0\\
};
\addlegendentry{SF$_{\rm min}=9$}

\addplot [semithick, red, mark=triangle, mark size=2, mark options={solid}, dashdotdotted]
table[row sep=crcr]{%
	0	0\\
};
\addlegendentry{SF$_{\rm min}=11$}


\end{axis}
\end{tikzpicture}%
    \end{subfigure}
    \vskip 0.1cm
    \centering 
    \subfloat[Average success probability, with $\rho_{\rm ID}= 50~\rm{ID/km}^2$.]
    {
        \label{fig:off_density}
\begin{tikzpicture}

\definecolor{chocolate2168224}{RGB}{216,82,24}
\definecolor{darkcyan0113188}{RGB}{0,113,188}
\definecolor{darkgray176}{RGB}{176,176,176}
\definecolor{goldenrod23617631}{RGB}{236,176,31}
\definecolor{olivedrab11817147}{RGB}{118,171,47}
\definecolor{purple12546141}{RGB}{125,46,141}

\begin{axis}[
width = \textwidth/2.3,
height = 5cm,
tick align=outside,
tick pos=left,
x grid style={darkgray176},
xmajorgrids,
xlabel={Density of TGs ($\rho_{\rm TG}$) [TG/km$^2$]},
xmin=0.1, xmax=0.5,
xtick style={color=black},
y grid style={darkgray176},
ymajorgrids,
ylabel={Average success probability $\left( \overline{{P}_S}\right)$},
ymin=0.52404, ymax=0.880315777460567,
ytick style={color=black}
]
\addplot [semithick, black, mark=square, mark size=2, mark options={solid}]
table {%
0.1 0.52404
0.2 0.72208
0.3 0.7976
0.4 0.8508
0.5 0.87828
};
\addplot [semithick, blue, dashed, mark=star, mark size=2, mark options={solid}, dashdotdotted]
table {%
0.1 0.683534201032654
0.2 0.759499696561355
0.3 0.816658045416686
0.4 0.854135450189671
0.5 0.877506728703578
};
\addplot [semithick, green, mark=diamond, mark size=2, mark options={solid}, dashdotdotted]
table {%
0.1 0.609480204453037
0.2 0.740027286773536
0.3 0.8064611386711
0.4 0.851680888784875
0.5 0.877803323966874
};
\addplot [semithick, red, mark=triangle, mark size=2, mark options={solid}, dashdotdotted]
table {%
0.1 0.552959384806953
0.2 0.730799782456123
0.3 0.807108007428268
0.4 0.849476829794497
0.5 0.879318189248538
};
\end{axis}

\end{tikzpicture}
    }\\\vspace{0.3cm}
     \subfloat[Average success probability, with $\rho_{\rm TG}= 0.1~\rm{TG/km}^2$.]
     {
        \label{fig:off_device}
\begin{tikzpicture}

\definecolor{chocolate2168224}{RGB}{216,82,24}
\definecolor{darkcyan0113188}{RGB}{0,113,188}
\definecolor{darkgray176}{RGB}{176,176,176}
\definecolor{goldenrod23617631}{RGB}{236,176,31}
\definecolor{olivedrab11817147}{RGB}{118,171,47}
\definecolor{purple12546141}{RGB}{125,46,141}

\begin{axis}[
width = \textwidth/2.3,
height = 5cm,
tick align=outside,
tick pos=left,
x grid style={darkgray176},
xmin=10, xmax=100,
xlabel={Density of IDs ($\rho_{\rm ID}$) [ID/km$^2$]},
xmajorgrids,
xtick style={color=black},
y grid style={darkgray176},
ylabel={Average success probability $\left( \overline{{P}_S}\right)$},
ymajorgrids,
ymin=0.42104, ymax=0.835921037075656,
ytick style={color=black}
]
\addplot [semithick, black, mark=square, mark size=2, mark options={solid}]
table {%
10 0.68708
25 0.60576
50 0.5056
75 0.46336
100 0.42104
};
\addplot [semithick, blue, dashed, mark=star, mark size=2, mark options={solid}, dashdotdotted]
table {%
10 0.775188255228844
25 0.737761454331635
50 0.684746069512848
75 0.635857869912958
100 0.596835389440926
};
\addplot [semithick, green, mark=diamond, mark size=2, mark options={solid}, dashdotdotted]
table {%
10 0.741656051447259
25 0.690322481640057
50 0.603519204895809
75 0.552440575482396
100 0.498378665927292
};
\addplot [semithick, red, mark=triangle, mark size=2, mark options={solid}, dashdotdotted]
table {%
10 0.704438521745232
25 0.638191081817121
50 0.552202770330313
75 0.486902871056561
100 0.435364209789591
};
\end{axis}

\end{tikzpicture}    
     }
     \caption{Average success probability vs. the TG density (left) and the ID density (right), considering different offloading options. We set $r=5$ km.}
\end{figure}

\subsection{Offloading}
\label{sub:res-opt}

In this section we consider a scenario in which IDs can offload data to a LEO satellite in the attempt to maximize the success probability, based on the optimization framework described in~\cref{sec_offloading}.
IDs (TGs) are uniformly distributed with density $\rho_{\rm ID}$ ($\rho_{\rm TG}$) over an AoI of radius $r=5$~km.
The payload size is fixed to 50 bytes, and the transmission rate is $\lambda= 1/360$ transmissions/s~\cite{3gpp}. 
Simulation results are given for SF$_{\rm min}=\{7,9,11\}$, i.e., as a function of the quality of the ID-L link as explained in~\cref{sec_offloading}), and benchmarked against a ``Standalone TG'' scheme in which data are processed onboard the TGs (i.e., $\eta^*=0$).



\paragraph{TG density}
In~\cref{fig:off_density} we evaluate the impact of the TG density in terms of the success probability, when $\rho_{\rm ID}=$ 50~ID/km$^2$. 
As expected, when the TG density is low, LEO offloading can increase the success probability by up to $+30\%$ compared to the ``Standalone TG'' baseline, especially in good channel conditions, i.e., when the received power in the ID-L is likely above the sensitivity threshold.
In particular, the additional computational capacity available at the LEO satellite can serve processing requests relative to cell-edge IDs, i.e., the most resource constrained network entities, which may otherwise not be able to communicate to TGs.
Moreover, when TGs are sparse, the ID-TG link is longer, which motivates more IDs to choose a higher SF to increase the coverage range, thus increasing the probability of collisions in the ``Standalone TG'' scenario. 
However, as the TG density increases, IDs are progressively closer to the TGs, and the more favorable channel on the ground 
gradually promotes onboard processing.



\paragraph{ID density}
In~\cref{fig:off_device} we study the success probability as a function of the ID density, when $\rho_{\rm TG}=$ 0.1~TG/km$^2$.
In general, as the ID density increases, the probability of collisions also increases, which may decrease the success probability to less than $50\%$ for $\rho_{\rm ID}=50$ ID/km$^2$ if ``Standalone TG'' is considered.
In turn, LEO offloading reduces the computational burden onboard the TGs, which improves the success probability, despite introducing some delays. 
Still, the benefit of the offloading in terms of success probability ranges from $+11\%$ when $\rho_{\rm ID}=10$ to around $+30\%$ when $\rho_{\rm ID}=50$ ID/km$^2$ in case of perfect channel conditions.

\section{Conclusions and Future Work}

\gls{ntn} is a promising technology to improve coverage and capacity of rural and remote areas. 
In particular, UAVs, HAPs, and satellites may serve as aerial/space gateways to collect and process IoT data from on-the-ground sensors, a paradigm referred to as NTN-IoT.
Along these lines, we evaluated the performance of different \gls{lpwan} technologies for IoT (i.e., LoRa, Sigfox, and NB-IoT) to communicate with \gls{ntn} platforms.
From our results, NB-IoT emerged as the most desirable technology to connect HAPs and UAVs, while LoRa turned out as the best approach for LEO satellites.
Based on that, we considered a scenario in which IoT sensors use LoRa to offload some data to \gls{leo} satellites, as a solution to alleviate the burden of data processing onboard the gateways. We demonstrate that LEO offloading can minimize the risk of collisions especially in sparsely-deployed networks, or when the density of sensors~increases.

As part of our future work, we will analyze the performance of the NTN-IoT paradigm considering the mobility of NTN platforms, and as a function of some other metrics such as energy consumption and latency.


\bibliographystyle{./bibliography/IEEEtran}
\bibliography{./bibliography/IEEEexample}

\end{document}